\documentclass[10pt,a4paper]{article} 
\usepackage{amssymb}
\usepackage{amsmath}
\usepackage{amsthm}
\usepackage{latexsym} 
\usepackage[dvips]{epsfig}
\usepackage{mathrsfs}
\usepackage{bm}

\theoremstyle{plain}
 
\newtheorem{lemma}{Lemma}
\newtheorem{theorem}{Theorem}

\newtheorem{conjecture}{Conjecture}

\newtheorem*{main}{Theorem}
\newtheorem*{definition}{Definition}

\font\SYM=msbm10 
\newcommand{\Real}{\mbox{\SYM R}}
\newcommand{\Complex}{\mbox{\SYM C}}

\newcommand{\Sphere}{\mbox{\SYM S}}

\font\tenscr=rsfs10 scaled1100
\font\sevenscr=rsfs7 
\font\fivescr=rsfs5 
\skewchar\tenscr='177 
\skewchar\sevenscr='177
\skewchar\fivescr='177 
\newfam\scrfam 
\textfont\scrfam=\tenscr
\scriptfont\scrfam=\sevenscr 
\scriptscriptfont\scrfam=\fivescr

\def\scri{{\fam\scrfam I}}

\renewcommand{\t}[1]{\tilde{#1}}
\renewcommand{\c}[1]{\cal{#1}}
\newcommand{\tc}[1]{\tilde{\cal{#1}}}

\begin{document}

\title{\textbf{Conformal extensions for stationary spacetimes}}

\author{{\Large Andr\'es E. Ace\~na} \thanks{E-mail address:
{\tt acena@aei.mpg.de}} \\
Max Planck Institut f\"ur Gravitationsphysik,\\
Albert Einstein Institut, \\
Am M\"uhlenberg 1, Golm,
D-14476 Germany.
\vspace{5mm}
\\
{\Large Juan A. Valiente Kroon} \thanks{E-mail address:
{\tt j.a.valiente-kroon@qmul.ac.uk}}\\
School of Mathematical Sciences, Queen Mary, University of London, \\
Mile End Road, London E1 4NS, UK.}

\date{March 2, 2011}

\maketitle

\begin{abstract}
The construction of the cylinder at spatial infinity for stationary
spacetimes is considered. Using a specific conformal gauge and frame,
it is shown that the tensorial fields associated to the conformal
Einstein field equations admit expansions in a neighbourhood of the
cylinder at spatial infinity which are analytic with
respect to some suitable time, radial and angular coordinates. It is
then shown that the essentials of the construction are independent of
the choice of conformal gauge. As a consequence, one finds that the
construction of the cylinder at spatial infinity and the regular
finite initial value problem for stationary initial data sets are, in
a precise sense, as
regular as they could be.
\end{abstract}


\section{Introduction}

The present article discusses a certain conformal extension for vacuum
stationary solutions for the Einstein field equations ---the so-called
\emph{cylinder at spatial infinity}. This construction provides
detailed information about the structure of this class of spacetimes in
the region where spatial infinity touches null infinity. The analysis
presented here is key for the construction of asymptotically simple
spacetimes from initial value problems on Cauchy hypersurfaces. In
order to better understand the context of our analysis, we present a
brief overview of the ideas and problems involved.

\subsection{Asymptotically simple spacetimes}
Penrose's notion of \emph{asymptotic simplicity} ---see
e.g. \cite{Pen63,PenRin86} was introduced with the objective of
providing a framework for the discussion of isolated systems in
General Relativity. The programme behind this idea is usually know as
\emph{Penrose's Proposal} ---see e.g. \cite{Fri99,Fri03a,Fri04}. A vacuum spacetime
$(\tilde{\mathcal{M}},\tilde{g}_{\mu\nu})$ is said to be
asymptotically simple if there exists a smooth, oriented, time-oriented,
causal spacetime $(\mathcal{M},g_{\mu\nu})$ and a smooth function
$\Xi$ (the \emph{conformal factor}) on $\mathcal{M}$ such that:
\begin{itemize}
\item[(i)] $\mathcal{M}$ is a manifold with boundary
$\mathscr{I}\equiv \partial \mathcal{M}$;
\item[(ii)] $\Theta>0$ on $\mathcal{M}\setminus \mathscr{I}$, and
$\Theta=0$, $\mbox{d} \Theta\neq 0$ on $\mathscr{I}$;
\item[(iii)] there exists an embedding 
$\Phi:\mathcal{M}\setminus \mathscr{I}\rightarrow \tilde{\mathcal{M}}$
such that $g_{\mu\nu}=\Theta^2(\Phi^{-1})^*\tilde{g}_{\mu\nu}$;
\item[(iv)] each null geodesic of $(\tilde{\mathcal{M}},\tilde{g}_{\mu\nu})$
acquires two distinct endpoints on $\mathscr{I}$.
\end{itemize}
In this definition, as well as in the rest of the article the word
\emph{smooth} will be used as synonym of $C^\infty$. In order to
simplify the notation we will write $g_{\mu\nu}=\Theta^2
\tilde{g}_{\mu\nu}$ instead of $g_{\mu\nu}=\Theta^2(\Phi^{-1})^*\tilde{g}_{\mu\nu}$. The
point (iv) in the definition excludes black hole spacetimes
---however, the 
discussion in this article will be local to the conformal boundary
$\mathscr{I}$, and hence (iv) will not be of relevance in our
considerations.

\medskip
The first natural examples of asymptotically simple spacetimes are
solutions to the Einstein field equations which are \emph{static} or,
more generally, \emph{stationary} near the conformal boundary. That
this is the case is a consequence of the structural properties of the
static and stationary field equations ---see e.g. \cite{Dai01b}. We
shall elaborate further on this in the coming paragraphs. Now, in
order to be of real physical value, the notion of asymptotic
simplicity should also include spacetimes which are not static or
stationary near the conformal boundary so as to allow for the
discussion of gravitational radiation ---the existence of this  type of spacetimes
has been shown in \cite{ChrDel02}. The examples constructed in
\cite{ChrDel02} are somehow special, as they arise as the
development of Cauchy initial data sets which are
exactly Schwarzschildean in the asymptotic end. More generally, recent
developments in the construction of solutions to the Einstein
constraint equations by means of gluing methods ---see
e.g. \cite{ChrDel03,Cor00,CorSch06}--- allow to construct asymptotically
simple spacetimes from initial data sets which are exactly stationary
in the asymptotic region. 

\medskip
Given this state of affairs, it is natural to ask whether there are
more general types of asymptotically simple spacetimes than the ones
described in the previous paragraph ---see e.g. \cite{Fri04}. This
question leads to the so-called \emph{problem of spatial infinity}. If asymptotically simple
spacetimes are to be constructed using some version of the Cauchy
problem in general Relativity, then one has to prescribe some initial
data set $(\tilde{\mathcal{S}},\tilde{h}_{ab},\tilde{\chi}_{ab})$ on
an \emph{asymptotically Euclidean hypersurface} ---for simplicity,
here it will be assumed that $\tilde{\mathcal{S}}$ has the topology of
$\Real^3$. The question now is: how does one encode in
$(\tilde{\mathcal{S}},\tilde{h}_{ab},\tilde{\chi}_{ab})$ that the 
development will be asymptotically simple? The examples of
\cite{ChrDel02} suggest that this issue has to be related in some way
to the behaviour of the initial data set in its asymptotic region. 

\subsection{Asymptotically Euclidean and regular initial data sets}
\label{Subsection:AsymptoticallyEuclidean}
As we are discussing properties of spacetimes by means of the
conformally rescaled setting given by the notion of asymptotic
simplicity, it is also natural to work with a conformally
rescaled one-point compactification of the initial hypersurface
$\tilde{\mathcal{S}}$. To this end, we recall the notion of
asymptotically Euclidean and regular Riemannian manifolds. The pair
$(\tilde{\mathcal{S}},\tilde{h}_{ab})$ will be said to be
\emph{asymptotically Euclidean and regular} if there exists a
3-dimensional, orientable, smooth compact manifold
$(\mathcal{S},h_{ab})$, a point $i\in \mathcal{S}$, a diffeomorphism
$\phi:\mathcal{S}\setminus\{i\}\rightarrow \tilde{\mathcal{S}}$ and a function
$\Omega\in C^2(\mathcal{S})\cap C^\infty(\mathcal{S}\setminus\{i\})$ with the
properties
\begin{eqnarray*}
&& \Omega(i)=0, \quad D_a\Omega(i)=0, \quad D_a D_b \Omega(i) = -2 h_{ab}(i), \\
&& \Omega>0 \quad \mbox{on } \mathcal{S}\setminus\{i\}, \\ 
&& h_{ab} = \Omega ^2 \phi_* \tilde{h}_{ab}.
\end{eqnarray*} 
Again, in order to simplify the notation, the last condition will be written
as $h_{ab} =\Omega^2 \tilde{h}_{ab}$ so that $\mathcal{S}\setminus
\{i\}$ and $\tilde{\mathcal{S}}$ are identified. The key property of
asymptotically Euclidean and regular manifolds is that suitable
neighbourhoods of the point $i$ (the \emph{point at infinity}) are
mapped into the asymptotic end of $\tilde{\mathcal{S}}$. Thus, one can
use local differential geometry to discuss the asymptotic properties
of the initial data set $(\tilde{\mathcal{S}},\tilde{h}_{ab})$.

\subsection{Static and stationary spacetimes}
\label{Subsection:IntroStationary}
The notion of asymptotically Euclidean and regular manifolds has been
crucial to understand the asymptotic properties of static and
stationary spacetimes ---see e.g. \cite{BeiSim80,Bei91b,Fri88}--- and
to prove results concerning their multipolar structure ---see
e.g. \cite{Ace09,BaeHer05a,Baeher05b,Baeher06,Bae07,BeiSim81a,BeiSim81b,Fri07}. Stationary
(and static) spacetimes are best discussed in terms of a \emph{quotient
manifold} $\tilde{\mathcal{X}}$ obtained by identifying points on
$\tilde{\mathcal{M}}$ lying on the same orbit of the stationary
(static) Killing vector $\xi^\mu$. From this symmetry reduction
procedure one also obtains a metric $\tilde{\gamma}_{ab}$ for the
quotient manifold $\tilde{\mathcal{X}}$. As the stationary spacetime
arises as the development of an asymptotically Euclidean initial data
set $(\tilde{\mathcal{S}},\tilde{h}_{ab},\tilde{\chi}_{ab})$, the pair
$(\tilde{\mathcal{X}},\tilde{\gamma}_{ab})$ will also be
asymptotically Euclidean. Conversely, one can prescribe the leading
asymptotic behaviour of the initial data set
$(\tilde{\mathcal{S}},\tilde{h}_{ab},\tilde{\chi}_{ab})$ from
assumptions on the asymptotic behaviour of
$(\tilde{\mathcal{X}},\tilde{\gamma}_{ab})$. For example, one can
assume that $(\tilde{\mathcal{X}},\tilde{\gamma}_{ab})$ is
asymptotically Euclidean and regular and work on a point-compactified
manifold $\mathcal{X}$ and a conformally rescaled quotient metric
$\gamma_{ab}$ ---this is the assumption made, for example, in
\cite{BeiSim80}.

\medskip
One of the key results of the theory of stationary spacetimes is that
there exists coordinates in a suitably small neighbourhood of $i$ for
which $\gamma_{ab}$, $\Omega$ and the (rescaled) stationary potentials
are analytic. This analyticity in a neighbourhood of the point at
infinity of the compactified quotient manifold is the key to establish
that static and stationary spacetimes are asymptotically simple. For
static spacetimes, the analyticity on $\mathcal{X}$ is inherited by
the point compactification $\mathcal{S}$ of time symmetric slices, and
the conformal factor $\Omega$ of the conformally rescaled quotient
metric $\gamma_{ab}$ is used as conformal factor for the whole spacetime,
so that the spacetimes are asymptotically simple. The situation for stationary
spacetimes is more delicate: in this case the (analytic)
conformal quotient metric $\gamma_{ab}$ is no longer conformally
related to the conformal metric $h_{ab}$ of the $t$-constant
slices. Furthermore, $h_{ab}$ is no longer analytic, but only of class
$C^{2,\alpha}$. Notwithstanding, it is still possible to construct a
smooth conformal extension of the stationary
spacetime. This result shows that although asymptotic
simplicity is a property which can be naturally expected from stationary
spacetimes, the fact that it holds is a consequence of the structural
properties of the stationary equations at spatial infinity.

\subsection{The cylinder at spatial infinity}
\label{Subsection:IntroCylinder}
In order to answer the question of whether there exist asymptotically
simple spacetimes arising from initial data sets which are neither
static nor stationary in a neighbourhood of infinity, one needs a
framework that allows to resolve and disentangle the delicate
structure of spacetime in this region. Furthermore, as the strategy to
construct asymptotically simple spacetimes is to make use of the
Cauchy problem in General Relativity, one would like to be able to
formulate an initial value problem with data prescribed on the compact
manifold $\mathcal{S}$ for various conformal fields which would
directly render the conformally rescaled manifold $(\mathcal{M},
g_{\mu\nu})$. Appropriate tools for this construction are the
\emph{conformal Einstein field equations} ---see
e.g. \cite{Fri81b,Fri81a,Fri82}--- and extensions thereof ---see
\cite{Fri95,Fri98a,Fri03a,Fri04}. However, the representation of
spatial infinity as suggested by the point-compactification of the
initial hypersurface $\tilde{\mathcal{S}}$ presents us with
technical difficulties. The underlying reason is that for initial data
sets with non-vanishing ADM mass, spatial infinity is a
singular point of the conformal geometry ---see e.g. \cite{Fri88}. At
the level of the conformal field equations and the various fields they
govern, this singular behaviour of the conformal structure translates
into the divergence of the so-called \emph{rescaled Weyl tensor} at
$i$.

\medskip
In order to overcome the difficulties at spatial infinity that have
been described in the previous paragraph, a new conformal
representation of the region of spacetime near null and spatial
infinity was introduced in \cite{Fri98a}. This representation, based
on general properties of conformal structures, together with the
\emph{extended conformal field equations} allows to introduce a
\emph{regular finite initial value problem at spatial infinity} for
which both the equations and their initial data are regular at the
conformal boundary.

\medskip
Whereas the standard (Penrose) compactification of spacetimes
considers spatial infinity as a point, the approach used in
\cite{Fri98a,Fri04} represents spatial infinity as an extended set
with the topology of $[-1,1]\times \Sphere^2$. This \emph{cylinder at
spatial infinity} is obtained as follows: one starts with the standard
point-compactification $\mathcal{S}$ of an asymptotically Euclidean
initial data set $\tilde{\mathcal{S}}$. As in previous paragraphs,
$\mathcal{S}$ contains a special point $i$ representing the infinity
of $\tilde{\mathcal{S}}$. In a second stage, the point $i$ is blown up
to a 2-sphere. This blowing up is achieved by lifting a neighbourhood
$\mathcal{B}$ of $i$ to the bundle of orthonormal frames with
group $O(3)$ ---or equivalently to the bundle of space-spinors with
group $SU(2,\Complex)$. In a final step, one uses a congruence of
timelike \emph{conformal geodesics} to obtain a conformal analogue of
Gaussian coordinates in a spacetime neighbourhood of
$\mathcal{B}$. Timelike conformal geodesics are conformal
invariants which preserve their quality of being timelike under
conformal transformations. Conformal Gaussian systems based on these
curves provide a canonical conformal class of conformal factors for
the development of the initial data. Remarkably, these conformal
factors can be written entirely in terms of initial data
quantities. Hence, the location of the conformal boundary is known
\emph{a priori}. The conformal boundary rendered by this class of
canonical conformal factors contains a null infinity with the same
structure as in the case of the standard Penrose
compactification. Spatial infinity, however, extends in the time
dimension ---so that one can speak of the cylinder at spatial infinity
proper. Of crucial relevance are the \emph{critical sets} $\{\pm
1\}\times \Sphere^2$ ---the collection of points where null and
spatial infinity intersect. Null infinity and spatial infinity do not
intersect tangentially at the critical points. As a consequence, some
of the propagation equations implied by the conformal field equations
degenerate at the critical points. The analysis in \cite{Fri98a}
---see also \cite{Val04a,Val04d,Val04e,Val05a}--- has shown that, as a
result, the solutions to the conformal field equations develop
certain types of logarithmic singularities at the critical sets. These
singularities form an intrinsic part of the conformal structure and
cannot be regarded as a consequence of a bad gauge choice. The
hyperbolic nature of the conformal propagation equations suggests that
these singularities will propagate along null infinity, and thus, they
will have an effect on the regularity of the conformal boundary.

\medskip
The construction of the cylinder at spatial infinity bears some
relation to other approaches in the analysis of the structure of
spatial infinity. For example, the blow up of the point at infinity is
closely related to Geroch's idea of directional dependent tensors
---see \cite{Ger72a,Ger76}. This idea was latter retaken in the discussions
given in \cite{AshHan78,Som78}. The cylinder at infinity is 
closely related to the \emph{hyperboloid of spatial infinity}
also introduced in \cite{AshHan78}, and latter retaken by
\cite{Bei84,BeiSch82} in a first attempt to combine geometric and
partial differential equations
points of view to study the structure of spatial infinity.

\subsection{Static spacetimes and the cylinder at spatial infinity}
\label{Subsection:StaticCylinder}
In view that static and stationary spacetimes provide prime
examples of asymptotically simple spacetimes, one also expects the
associated 
construction of the cylinder at spatial infinity to be as smooth as it
can be. This smoothness can be regarded as a consistency of the
setting. If static or stationary spacetimes were to exhibit some type
of non-smooth behaviour at the cylinder at spatial infinity, these by
necessity have to be associated to a bad gauge choice.  In \cite{Fri04} a proof of the
smoothness of the cylinder at spatial infinity for static spacetimes
was given. Surprisingly, this proof is much more complicated than what
one would expect given that: firstly, the conformal fields are
analytic in a neighbourhood of spatial infinity; and secondly, that
the spacetimes are time independent. The difficulties in the analysis
can be explained, in part, by the fact that the conformal geodesics used
in the construction of the cylinder at spatial infinity are not
aligned with the orbits of the timelike Killing vector. Nevertheless,
the fact that one is considering time symmetric spacetimes simplifies
the analysis as the quotient manifold can be identified with the
slices of constant $t$ so that the analyticity of the point $i$
is inherited by spatial infinity $i^0$ and by all spacetime quantities.

\medskip
It should be pointed out that the relevance of the analysis of static spacetimes
carried out in \cite{Fri04} goes beyond its role as a consistency check of the
framework. The analysis in \cite{Val04a,Val04e} suggests that static
spacetimes have a special position among the class of spacetimes with
a smooth compactification at spatial infinity. More precisely, it is
conjectured that:

\begin{conjecture}
\label{ConjectureStatic}
If an analytic time symmetric initial data set for the Einstein vacuum
equation yields a development with a smooth null infinity, then the
initial data set is exactly static in the neighbourhood of spatial infinity.
\end{conjecture}

The rigidity results of \cite{Val10a,Val10b} constitute further
evidence in support of the conjecture.

\subsection{Our main result}
\label{Subsection:MainResult}
In the present article we extend the analysis of the cylinder at
spatial infinity carried out in \cite{Fri04} to the case of of
stationary spacetimes.  More precisely, we show that:

\begin{main}
Given an initial data set for the vacuum Einstein field equations
which is stationary in a neighbourhood of infinity, the solutions to
the regular finite initial value problem for the conformal field
equations at spatial infinity is smooth in a neighbourhood of the cylinder at spatial
infinity, and in particular through the critical sets.
\end{main}

The strategy to prove this result is as follows. Starting with a
generic asymptotically flat stationary spacetime
$(\tilde{\mathcal{M}},\tilde{g}_{\mu\nu})$, one makes use of the analysis of
\cite{Dai01b} to construct a conformal completion, $(\breve{\Omega},\breve{g}_{\mu\nu})$, of the stationary
spacetime in a neighbourhood of spatial infinity. This completion is
adapted to the stationarity of the spacetime, and has a smooth null
infinity. This representation is, however, not suitable for our
purposes as spatial infinity is represented as a point. The conformal
metric $\breve{g}_{\mu\nu}$ is given in terms of some asymptotically
Cartesian coordinates. In a second stage one performs a change of
coordinates to a polar system in which asymptotic expansions can be
analysed in a more convenient way. A subsequent change of the time
coordinate and an associated conformal rescaling render a conformal
representation, $(\bar{\Omega},\bar{g}_{\mu\nu})$, in which spatial
infinity already appears as an extended set with the topology of a
cylinder. This representation is, in a strict sense, not a conformal
completion as spatial infinity is not at a finite distance with
respect to the conformal metric $\bar{g}_{\mu\nu}$ ---the metric
becomes singular there. In order to deal with this behaviour one introduces
a suitable frame $v_{\bm \alpha}$. It is then show that the components
of the metric, $\bar{g}_{\bm \alpha \bm \beta}$,
with respect to this frame are regular at infinity. Furthermore, the
components of key derived objects (the Schouten and Weyl tensors) are
also shown to be regular. This is the most calculational involved
part of our argument. Once the cylinder at spatial infinity has been
obtained, one shows that the cylinder itself, and a neighbourhood of
it can be ruled by means of a congruence of conformal geodesics. This
cannot be shown explicitly, and thus, one has to resort to a
perturbative argument. This
construction leads to the \emph{canonical} conformal factor $\Theta$,
related to $\bar{\Omega}$ through a further conformal factor
$\Pi$. The congruence of conformal congruences also gives rise to a
Weyl connection $\hat{\nabla}$.  An abstract integration
of the conformal evolution equations along the congruence of conformal
geodesics shows that solutions to the conformal evolution equations
extend smoothly through the cylinder at spatial infinity and also
through a suitable neighbourhood of null infinity. In a final step, it
is then shown that the construction is independent of the conformal
gauge used to write the stationary initial data. 

\medskip
Contrasted with the result for static spacetimes given in
\cite{Fri04}, the main difficulties in proving our main result are:
the presence of a non-vanishing second fundamental form in the slices
of  constant $t$ has as a consequence a Weyl tensor with
non-vanishing magnetic part; and crucially, the quotient manifold cannot be
directly identified with the structures of the constant $t$
slices. In particular, as already discussed, the analytic structure of the quotient
manifold in a neighbourhood of infinity is not inherited by the
slices of constant coordinate $t$. Instead, one obtains fields which are of
the form $f+ \rho g$ with $f$, $g$ analytic and $\rho$ a suitable radial
coordinate ---recall that the radial coordinate is not smooth in a neighbourhood
of $i$ with respect to Cartesian coordinates. It is of interest to
notice that our analysis requires the explicit computation up to
quadrupolar order of the expansions of the relevant conformal fields.

\medskip
Our argument assumes, \emph{a priori},
the existence of the stationary spacetime, and makes statements about
the smoothness of the spacetime in a certain gauge from the known
smoothness in another gauge. A proof that makes only use of the
conformal evolution and properties of stationary initial data sets
would be much more complicated and would require a much deeper
understanding of the properties of the conformal field
equations and associated conformal structures than the one that is
currently available. We expect our analysis to shed some light in this direction.

\medskip
As in the case of static spacetimes, our main result, on the one hand,
ensures that the construction of the cylinder at spatial infinity for
spacetimes without time reflexion symmetry does not have spurious
gauge singularities, and on the other hand, it is expected to play a
key role in a proof of a suitable generalisation of Conjecture \ref{ConjectureStatic}.

\subsection*{Overview of the article}
In Section \ref{fieldEquations} we present a concise discussion of the
Conformal field equations and conformal geodesics. The presentation in
this section is aimed at providing a context for the analysis of the
subsequent sections of this article. Section \ref{regularProblem}
briefly summarises the so-called regular initial value problem at
spatial infinity. This discussion includes, in particular, the
construction of the so-called cylinder at spatial infinity. Section
\ref{stationarySection} discusses results about stationary spacetimes
which are relevant for our analysis. Particular attention is paid to
their asymptotic expansions in both the quotient manifold and in a
Cauchy slice. A ``standard'' conformal completion of stationary
spacetimes is discussed. Section \ref{confExtSVS} discusses an
alternative conformal completion for stationary spacetimes. This
particular completion ultimately leads to the cylinder at spatial
infinity. A discussion of the asymptotic expansions for the relevant
field quantities in this conformal completion are provided. In
particular, it is shown that the components of the Schouten and Weyl
tensors with respect to a particular frame are regular at
infinity. Section \ref{Section:CylinderStationary} provides a
discussion of the construction of conformal Gaussian systems in the
neighbourhood of the cylinder at spatial infinity of stationary
spacetimes. As a result of this analysis, it is shown that in a
certain gauge the setting of the initial value problem at spatial
infinity is as regular as it is to be expected. In a second stage it
is shown that the construction is independent of the particulars of
the choice of conformal gauge. This discussion completes the proof of
our main result. Section \ref{Section:Conclusions} provides some
concluding remarks concerning our analysis. Some lengthy expansions
are presented separately from the main text in Appendix \ref{expansions}.

\section{Conformal field equations and conformal geodesics}\label{fieldEquations}

The \textit{regular finite initial value problem at spatial infinity},
presented in Section \ref{regularProblem}, was introduced by Friedrich
in \cite{Fri98a} and is based on a conformal representation of the Einstein
field equations, known as the \textit{extended conformal field
equations}. In this section  we elaborate further on the ideas
discussed in Subsection \ref{Subsection:IntroCylinder} of the
introduction, and we present  a concise discussion of this
system and of its associated structures. The presentation is geared
towards the purposes of the present article.

\subsection{Weyl connections}
Let $\tc{M}$ denote a 4-dimensional manifold endowed with a Lorentzian
metric $\t{g}_{\mu\nu}$. A conformal rescaling of the metric is given by
\[
 \t{g}_{\mu\nu}\rightarrow g_{\mu\nu}=\Theta^2\t{g}_{\mu\nu}
\]
where $\Theta$ is a positive function on $\tc{M}$. The conformal
class $[\t{g}]$ is the collection of all metrics conformally
related to $\t{g}_{\mu\nu}$
\[
[\t{g}]\equiv\{ g_{\mu\nu}\, |\, g_{\mu\nu}=\Theta^2\tilde{g}_{\mu\nu},\,\Theta>0\}.
\]
Let $\hat{\nabla}$ denote the covariant derivative operator of a
torsion-free connection on $\tc{M}$. This connection is called a
\emph{conformal connection} or \emph{Weyl connection} for $[\t{g}]$ if
for $g_{\mu\nu}\in[\t{g}]$ one has that
\begin{equation}\label{weylDer} \hat{\nabla}_\mu
g_{\nu\lambda}=-2\,f_\mu\ g_{\nu\lambda}
\end{equation} for some smooth 1-form $f_\mu$. The connection
$\hat{\nabla}$ preserves the conformal structure of
$[\tilde{g}]$. Furthermore, it does not depend on the class
representative. For example, if
$\check{g}_{\mu\nu}=\check{\Theta}^2g_{\mu\nu}$, then
\[
 \hat{\nabla}_\mu \check{g}_{\nu\lambda}=-2\,\check{f}_\mu\, \check{g}_{\nu\lambda},
\]
with
\[
 \check{f}_\mu=f_\mu-\check{\Theta}^{-1}\partial_\mu\check{\Theta}.
\]
If $\nabla$ is the Levi-Civita connection of $g_{\mu\nu}$, then from
equation \eqref{weylDer} we have that the connections $\nabla$ and
$\hat{\nabla}$ define the difference tensor
$\hat{\nabla}-\nabla=S(f)$, given by
\begin{equation}\label{diffTensor}
 S(f)_\mu\,^\rho\,_\nu=\delta_\mu{}^\rho\, f_\nu+\delta_\nu{}^\rho\, f_\mu-g_{\mu\nu}g^{\rho\lambda}f_\lambda,
\end{equation}
where $\delta_\mu{}^\rho$ is the Kronecker delta. This shows that $\hat{\nabla}$ can be specified using $\nabla$ and $f_\mu$.

\subsection{The extended conformal field equations}
We now specialise to the case where $\t{g}_{\mu\nu}$ is a solution of
Einstein vacuum field equations on $\tc{M}$, 
$\mbox{Ric}[\t{g}_{\mu\nu}]=0$. The Weyl connection
$\hat{\nabla}$  and the Levi-Civita connections $\nabla$ and $\t{\nabla}$ are related by
\[
 \hat{\nabla}-\t{\nabla}=S(\t{f}),\hspace{1cm}\nabla-\t{\nabla}=S(\Theta^{-1}\mbox{d}\Theta),\hspace{1cm}\hat{\nabla}-\nabla=S(f).
\]
The Riemann tensor of  the Weyl connection $\hat{\nabla}$ can be
decomposed as 
\[
 \hat{R}^\mu\,_{\nu\lambda\rho}=2(g^\mu\,_{[\lambda}\hat{L}_{\rho]\nu}-g^\mu\,_\nu\hat{L}_{[\lambda\rho]}-g_{\nu[\lambda}\hat{L}_{\rho]}\,^\mu)+C^\mu\,_{\nu\lambda\rho},
\]
where $\hat{L}_{\mu\nu} $ and $C^\mu\,_{\nu\lambda\rho}$ denote,
respectively, the Schouten and Weyl tensors. The Schouten
tensor  is given by
\[
 \hat{L}_{\mu\nu}=\frac{1}{2}\hat{R}_{(\mu\nu)}-\frac{1}{4}\hat{R}_{[\mu\nu]}-\frac{1}{12}\hat{R}\,g_{\mu\nu}
\]
where
\[
\hat{R}_{\mu\nu}=\hat{R}^\rho\,_{\mu\rho\nu}, \quad  \hat{R}=g^{\mu\nu}\hat{R}_{\mu\nu}.
\]

\medskip
In the sequel it will be convenient to consider the $1$-form
\[
 d_\mu\equiv\Theta \t{f}_\mu=\Theta f_\mu+\nabla_\mu\Theta,
\]
and the rescaled conformal Weyl tensor
\[
 W^\mu\,_{\nu\lambda\rho}=\Theta^{-1}C^\mu\,_{\nu\lambda\rho}.
\]

\medskip
In order to deal with the possible direction
dependence of the various fields near space-like infinity, it is convenient to use a frame formalism and a suitably chosen
orthonormal frame field. For this,
let us take a frame field $\{e_{\bm\alpha}\}_{{\bm\alpha}={\bm0},{\bm1},{\bm2},{\bm3}}$ satisfying
\[g_{\bm{\alpha\beta}}\equiv
g(e_{\bm{\alpha}},e_{\bm{\beta}})=\eta_{\bm{\alpha\beta}}=\mbox{diag}(+1,-1,-1,-1).
\]
Let $\hat{\nabla}_{\bm\alpha}$ and $\nabla_{\bm\alpha}$ denote, respectively, the
covariant derivatives in the direction of $e_{\bm\alpha}$ with respect to
the connection $\hat{\nabla}$ and $\nabla$. The
respective connection coefficients are defined by $\hat{\nabla}_{\bm\alpha}
e_{\bm\beta}=\hat{\Gamma}_{\bm\alpha}{}^{\bm\gamma}{}_{\bm\beta}\,e_{\bm\gamma}$ and
$\nabla_{\bm\alpha} e_{\bm\beta}=\Gamma_{\bm\alpha}{}^{\bm\gamma}{}_{\bm\beta}\,e_{\bm\gamma}$. The
frame coefficients with respect to an as yet unspecified coordinate
system $x^\mu$ are given by $e^\mu{}_{\bm\alpha}=\langle
\mbox{d}x^\mu,e_{\bm\alpha}\rangle$. Using equation \eqref{diffTensor} we have
\[
\hat{\Gamma}_{\bm\alpha}{}^{\bm\gamma}{}_{\bm\beta}=\Gamma_{\bm\alpha}{}^{\bm\gamma}{}_{\bm\beta}+\delta_{\bm\alpha}{}^{\bm\gamma}\,f_{\bm\beta}+\delta_{\bm\beta}{}^{\bm\gamma}\,f_{\bm\alpha}-g_{\bm{\alpha\beta}}g^{\bm{\gamma\delta}}f_{\bm\delta},
\]
where $f_{\bm\alpha}=e^\mu{}_{\bm\alpha} f_\mu$. Then, if all tensor fields are
expressed in terms of the frame field and the corresponding connection
coefficients, the extended conformal field equations are equations for the
unknowns
\[
u\equiv (e^\mu{}_{\bm\alpha},\,\hat{\Gamma}_{\bm\alpha}{}^{\bm\gamma}{}_{\bm\beta},\,\hat{L}_{\bm{\alpha\beta}},\,W^{\bm\alpha}{}_{\bm{\beta\gamma\delta}}),
\]
and are given by
\begin{subequations}
\begin{eqnarray}
 &&
 [e_{\bm\alpha},e_{\bm\beta}]=(\hat{\Gamma}_{\bm\alpha}{}^{\bm\gamma}{}_{\bm\beta}-\hat{\Gamma}_{\bm\beta}{}^{\bm\gamma}{}_{\bm\alpha})e_{\bm\gamma}, \label{eqConfFrame1}\\
&&
e_{\bm\alpha}(\hat{\Gamma}_{\bm\beta}{}^{\bm\delta}{}_{\bm\gamma})-e_{\bm\beta}(\hat{\Gamma}_{\bm\alpha}{}^{\bm\delta}{}_{\bm\gamma})-(\hat{\Gamma}_{\bm\alpha}{}^{\bm\epsilon}{}_{\bm\beta}-\hat{\Gamma}_{\bm\beta}{}^{\bm\epsilon}{}_{\bm\alpha})\hat{\Gamma}_{\bm\epsilon}{}^{\bm\delta}{}_{\bm\gamma}+\hat{\Gamma}_{\bm\alpha}{}^{\bm\delta}{}_{\bm\epsilon}\hat{\Gamma}_{\bm\beta}{}^{\bm\epsilon}{}_{\bm\gamma}-\hat{\Gamma}_{\bm\beta}{}^{\bm\delta}{}_{\bm\epsilon}\hat{\Gamma}_{\bm\alpha}{}^{\bm\epsilon}{}_{\bm\gamma} \nonumber
\\
&& \hspace{2cm} =2\{g^{\bm\delta}{}_{[\bm\alpha}\hat{L}_{\bm\beta]\bm\gamma}-g^{\bm\delta}{}_{\bm\gamma}\hat{L}_{[\bm\alpha\bm\beta]}-g_{\bm\gamma[\bm\alpha}\hat{L}_{\bm\beta]}\,^{\bm\delta}\}+\Theta
W^{\bm\delta}{}_{\bm\gamma\bm\alpha\bm\beta},  \label{eqConfFrame2} \\
&&
 \hat{\nabla}_{\bm\alpha} \hat{L}_{\bm\beta\bm\gamma}-\hat{\nabla}_{\bm\beta} \hat{L}_{\bm\alpha\bm\gamma}=d_{\bm\delta} W^{\bm\delta}{}_{\bm\gamma\bm\alpha\bm\beta}, \label{eqConfFrame3}\\
&& \nabla_{\bm\delta} W^{\bm\delta}{}_{\bm\gamma\bm\alpha\bm\beta}=0. \label{eqConfFrame4}
\end{eqnarray}
\end{subequations}
In the last equation, called the \textit{Bianchi equation}, one has to
consider the relation between $\hat{\Gamma}_{\bm\alpha}{}^{\bm\gamma}{}_{\bm\beta}$
and $\Gamma_{\bm\alpha}{}^{\bm\gamma}{}_{\bm\beta}$, whence
$f_{\bm\alpha}=\frac{1}{4}\hat{\Gamma}_{\bm\alpha}{}^{\bm\beta}{}_{\bm\beta}$. Notice that
no differential equations are given for the fields $\Theta$ and
$d_{\bm\alpha}$. This is due to the conformal gauge freedom introduced into
Einstein's field equations by considering general Weyl connections and
conformal metrics.

\subsection{The conformal Gauss gauge}
The fields $\Theta$ and $d_{\bm\alpha}$ can be specified by means of a
choice of gauge. To this end we consider conformal
geodesics. A conformal geodesic for $(\tc{M},\t{g}_{\mu\nu})$ is a curve
$x(\tau)$ in $\tc{M}$ and a $1$-form $\t{f}(\tau)$ along the curve,
which solve the following system of ordinary differential equations:
\begin{subequations}
\begin{eqnarray}
&&
(\tilde{\nabla}_{\dot{x}}\dot{x})^\mu+S(\t{f})_\lambda{}^\mu{}_\rho
\,\dot{x}^\lambda\,\dot{x}^\rho=0, \label{geod1} \\
&& 
 (\tilde{\nabla}_{\dot{x}}\t{f})_\nu-\tfrac{1}{2}\t{f}_\mu\,
 S(\t{f})_\lambda{}^\mu{}_\nu\,\dot{x}^\lambda=\tilde{L}_{\lambda\nu}\,\dot{x}^\lambda. \label{geod2}
\end{eqnarray}
\end{subequations}
Conformal geodesics are invariants of the conformal structure in the
following sense: if $x(\tau)$, $\t{f}(\tau)$ solve equations
\eqref{geod1}-\eqref{geod2} and $b$ is a smooth 1-form field on
$\tc{M}$, then the pair $x(\tau)$, $(\t{f}-b)|_{x(\tau)}$ solves
the same equations with $\t{\nabla}$ replaced by
$\hat{\nabla}=\t{\nabla}+S(\t{b})$ and $\t{L}$ by $\hat{L}$. This
means that $x(\tau)$, and in particular the parameter $\tau$, do not
depend on the Weyl connection in the conformal class which is used to
write the conformal geodesic equations.

\medskip
Conformal geodesics, and in particular congruences of conformal
geodesics, can be used to construct a special gauge for the conformal
equations. For this, let $\tc{S}$ be a space-like hypersurface in the
given vacuum solution $(\tc{M},\t{g}_{\mu\nu})$. We choose on $\tc{S}$
a positive `conformal factor' $\Theta_*$, a frame field $e_{\bm\alpha*}$,
and a $1$-form $\t{f}_*$, such that
$\t{g}(e_{\bm\alpha*},e_{\bm\beta*})=\Theta_*^{-2}\eta_{\bm\alpha\bm\beta}$ and
$e_{\bm0*}$ is orthogonal to $\tc{S}$. Then there exists through each
point $x_*\in\tc{S}$ a unique conformal geodesic $x(\tau)$,
$\t{f}(\tau)$ with $\tau=0$ on $\tc{S}$ which satisfies there the
initial conditions $\dot{x}=e_{\bm0*}$, $\t{f}=\t{f}_*$. These curves
define a smooth caustic free congruence in a neighbourhood $\c{U}$ of
$\tc{S}$ if all data are smooth. Furthermore, $\t{f}$ defines a smooth
$1$-form on $\c{U}$ which supplies a Weyl connection $\hat{\nabla}$ on
$\c{U}$ given by $\hat{\nabla}=\t{\nabla}+S(\t{f})$. A smooth frame
field $e_{\bm\alpha}$ and a conformal factor $\Theta$ are then obtained on
$\c{U}$ by solving
\begin{eqnarray*}
&& \hat{\nabla}_{\dot{x}}e_{\bm\alpha}=0, \\
&& \hat{\nabla}_{\dot{x}}\Theta=\Theta\,\langle\dot{x},\t{f}\rangle,
\end{eqnarray*}
for given initial conditions $e_{\bm\alpha}=e_{\bm\alpha*}$, $\Theta=\Theta_*$
on $\tc{S}$. The frame field is orthonormal for the metric
$g_{\mu\nu}=\Theta^2\tilde{g}_{\mu\nu}$. Dragging along the congruence the local
coordinates $x^a$, $a=1,2,3$ on $\tc{S}$ and setting $x^0=\tau$ we
obtain a coordinate system. In this gauge one has in $\c{U}$ that 
\begin{equation}\label{gaugeFrame}
 \dot{x}=e_{\bm0}=\partial_\tau,\quad
 \hat{\Gamma}_{\bm0}{}^{\bm\beta}{}_{\bm\alpha}=0, \quad \hat{L}_{\bm0\bm\alpha}=0.
\end{equation}
This choice of coordinates, frame field, and conformal gauge will be
referred to as {\it conformal Gauss system}. Remarkably, in this gauge
it is possible to obtain explicit expressions for $\Theta$ and $d$ in terms of the initial data,
given by
\begin{subequations}
\begin{eqnarray}
 &&
 \Theta(\tau)=\Theta_*\Big(1+\langle\t{f}_*,e_{{\bm 0}*}\rangle\tau+\frac{1}{4}g^\sharp(\t{f}_*,\t{f}_*)\tau^2\Big), \label{confFact}
 \\
&&
 d_{\bm 0}=\dot{\Theta},\hspace{2cm}d_{\bm
   a}=\Theta_*\langle\t{f}_*,e_{\bm a*}\rangle,\hspace{1cm}{\bm a}=1,2,3, \label{compb}
\end{eqnarray}
\end{subequations}
where the quantities with the subscript `$*$' are constant along the
conformal geodesics and $g^\sharp$ denotes the contravariant version
of $g_{\mu\nu}$. These expressions, together with equations \eqref{eqConfFrame1}-\eqref{eqConfFrame4}
provide a complete system of equations for $u$, called the {\it
general conformal field equations}. Setting
$\bm \alpha= 0$ in \eqref{eqConfFrame1}-\eqref{eqConfFrame4} and observing
the gauge conditions \eqref{gaugeFrame} one obtains the following
evolution equations:
\begin{subequations}
\begin{eqnarray}
 && \partial_\tau
 e^\mu{}_{\bm\alpha}=-\hat{\Gamma}_{\bm\alpha}{}^{\bm\beta}{}_{\bm 0} e^\mu{}_{\bm\beta}, \label{redEq1} \\
&&
 \partial_\tau\hat{\Gamma}_{\bm\alpha}{}^{\bm\gamma}{}_{\bm\beta}=-\hat{\Gamma}_{\bm\delta}{}^{\bm\gamma}{}_{\bm\beta}\hat{\Gamma}_{\bm\alpha}{}^{\bm\delta}{}_{\bm0}+g^{\bm
   \gamma}{}_{\bm0}\hat{L}_{\bm\alpha\bm\beta}+g^{\bm\gamma}{}_{\bm\beta}\hat{L}_{\bm0\bm\alpha}-g_{\bm\beta\bm0}\hat{L}_{\bm\alpha}{}^{\bm\gamma}+\Theta\,W^{\bm\gamma}{}_{\bm\beta\bm0\bm\alpha}, \label{redEq2} \\
&&
 \partial_\tau\hat{L}_{\bm{\alpha\beta}}=d_{\bm\gamma}
 W^{\bm\gamma}\,_{\bm{\alpha0\beta}} \label{redEq3}. 
\end{eqnarray}
\end{subequations}
If one extracts from the Bianchi equation \eqref{eqConfFrame4}
 a symmetric hyperbolic system, one gets symmetric hyperbolic reduced
equations for those components of $u$ which are not determined
explicitly by the gauge conditions. The resulting system is called the
\textit{reduced conformal field equations}. It can be shown that for
such a choice of reduced equations, any solution which satisfies
\eqref{eqConfFrame1}-\eqref{eqConfFrame4} on a suitable spacelike
hypersurface does indeed satisfy the complete set of field equations
in the part of the domain of dependence of the initial data set where
$\Theta$ is positive. This equations are equivalent to Einstein field
equations in the sense that if one has a solution of the conformal
system, then one has a solution of Einstein field equations in the
region where the conformal factor is positive, and \emph{vice versa}. The
substantial advantage of using the conformal system is that one can
deal with regions where the conformal factor vanish. Furthermore,
in the Gauss gauge the location of this region can be prescribed \emph{a
priori}, giving us full control on the conformal boundary if the
evolution extends far enough.

\section{The regular finite initial value problem at space-like infinity}\label{regularProblem}

As mentioned in the introduction, the construction of the regular
finite initial value problem at space-like infinity consists of two
main steps. For this, one considers a hypersurface $\c{S}$, which is a
one-point conformal compactification of the Cauchy hypersurface
$\tc{S}$ and therefore contains a geometrically distinguished point
$i$ ---cfr. the discussion in Subsection
\ref{Subsection:AsymptoticallyEuclidean}. In the
first step of the construction, the point $i$ is blown up into a
sphere ${\c{I}}^0$. In the second step, a congruence of conformal
geodesics is used to describe the evolution of the conformal fields in a
neighbourhood of ${\c{I}}^0$. The construction is described in detail
in \cite{Fri04}, and it is implemented through the bundle of spin
frames over $\c{S}$ near $i$. In what follows, we present a summary of
the aspects of a similar construction based on the bundle of
orthonormal frames  ---see also \cite{Fri03a}.

\subsection{The blow up of spatial infinity}
Start by considering a space-like Cauchy hypersurface
$\c{S}$ with intrinsic metric $h_{ab}$ containing the distinguished
point $i$. Next, choose a fixed oriented $h$-orthonormal frame $e_{\bm
  a}$,
${\bm a}=1,2,3$, at $i$. Any other such frame at $i$ is obtained by a
rotation of $e_{\bm a}$. That is, all other $h$-orthonormal frames at $i$
are of the form $e_{\bm a}(s)=s^{\bm b}{}_{\bm a}\,e_{\bm b}$ with
$s=(s^{\bm b}{}_{\bm a})\in
SO(3)$. In particular, $e_{\bm 3}(s)$ covers all possible directions at $i$
as one lets $s$ exhaust $SO(3)$. For a given value of $s$, one
distinguishes $e_{\bm 3}(s)$ as the radial vector at $i$. Keeping $s$ fixed,
we construct the $h$-geodesic starting at $i$ that has tangent vector
$e_{\bm 3}(s)$ and denote by $\rho$ the affine parameter on the geodesic
that vanishes at $i$. The frame $e_{\bm a}(s)$ is then parallelly
transported along the geodesic. For a particular value of the affine
parameter $\rho$, the frame thus obtained will be
denoted by $e_{\bm a}(\rho,s)$. We will consider only the region $|\rho|<a$,
where $a$ is chosen such that the metric ball $\c{B}$ centered at $i$
with radius $a$ is a convex normal neighbourhood for the 3-metric
$h$. The map
from the set $(-a,a)\times SO(3)$ into the bundle $SO(\c{S})$ of
oriented orthonormal frames over $\c{S}$, given by
$(\rho,s)\rightarrow e_{\bm a}(\rho,s)$, defines a smooth embedding of a
$4$-dimensional manifold into $SO(\c{S})$.

\medskip
In what follows, only non-negative values of $\rho$ will be
considered. Denote by $\hat{\c{B}}$ the image of the set $[0,a)\times
SO(3)$. The boundary of $\hat{\c{B}}$ will be denoted by
${\c{I}}^0$. One has that ${\c{I}}^0=\{\rho=0\}\simeq SO(3)$. Finally,
let $\pi$ denote the restriction to $\hat{\c{B}}$ of the projection of
$SO(\c{S})$ onto $\c{S}$. In the sequel it will be convenient to consider the subgroup $SO(2)$ of
$SO(3)$, given by $SO(2)=\{s'\in SO(3)\,|\,s'^{\bm b}{}_{\bm 3} e_{\bm
  b}=e_{\bm 3}\}$ ---this
is the subgroup of $SO(3)$ whose action leaves $e_{\bm 3}$
invariant. Accordingly, if  $s\in
SO(3)$ and $s'\in SO(2)$ then $e_{\bm a}(s)$ and $e_{\bm a}(s\,s')$ are parallelly transported along the
same geodesic. Hence, when we consider the projection $\pi$ we have that
$\pi(e_{\bm a}(\rho,s))=\pi(e_{\bm a}(\rho,s\,s'))$, and therefore the map $\pi$
has a factorisation
\[
\hat{\c{B}}\xrightarrow{\pi'}{\c{B}}'=\hat{\c{B}}/SO(2)\xrightarrow{\pi''}\c{B}.
\]
On the one hand, the projection $\pi''$ maps the set
$\pi'({\c{I}}^0)\simeq S^2$ onto $i$  and on the other 
it implies a diffeomorphism of ${\c{B}}'\backslash\pi'({\c{I}}^0)$
onto the punctured ball ${\c{B}}\backslash\{i\}$. This diffeomorphism can be used
to identify these sets. However, instead of this, it is convenient to
pull back the initial data on $\c{B}$ to $\hat{\c{B}}$ via $\pi$. Then
$\hat{\c{B}}$ becomes the initial manifold, $\rho$ and $s$ are used as
coordinates on it, and its boundary ${\c{I}}^0$ is a blow up of
$i$. This manifold has an extra dimension when compared to the initial
hypersurface $\c{B}$. This extra dimension is given by the action of
$SO(2)$. Since all fields have a well defined transformation behaviour
(spin weight) under such action, on the part of $\hat{\c{B}}$ where
$\rho>0$  vector fields $X$, $c_{\bm a}(\rho,s)$, ${\bm a}=1,2,3$, can be
prescribed such that $X$ is generated by the action
of $SO(2)$ and the vector fields $c_{\bm a}(\rho,s)$ satisfy
$T(\pi)c_{\bm a}(\rho,s)=e_{\bm a}(\rho,s)$ ---see \cite{Fri04}. These vector fields allow the
introduction of a frame formalism.

\subsection{Implementing the conformal Gauss gauge}
In the second step of the construction, the development of the data is
considered. For this, one uses a conformal Gauss system ---see 
Section \ref{fieldEquations}. In what follows, it is assumed that the conformal compactification
of the hypersurface $\tc{S}$ has been achieved by means of a conformal
factor $\Omega$. The initial data for the conformal factor for the Gauss
system is set by requiring that
\begin{equation}\label{Thetastar}
  \Theta_*\equiv\kappa^{-1}\Omega.
\end{equation}
where the function $\kappa$ satisfies
\[
 \kappa = \rho\kappa', \quad \kappa'\in C^\infty(\hat{\c{B}}),\quad
 \kappa'>0,\quad X\kappa'=0,\quad \kappa'|_{{\c{I}}^0}=1.
\]
The change of the conformal factor induced by the function $\kappa$
implies a map $\Xi:e_{\bm a}\rightarrow \kappa e_{\bm a}$ which maps the set
$\hat{\c{B}}\backslash{\c{I}}^0$ bijectively onto a smooth submanifold
$\c{B}^*$ of the bundle of frame fields over $\c{B}$. The
diffeomorphism $\Xi$ is used to carry the coordinates $\rho$, $s$ and
the vector fields $X$, $c_{\bm a}$ to $\c{B}^*$. The projection of
$\c{B}^*$ onto $\c{B}$ will be denoted again by $\pi$. 

\medskip
The construction of the conformal Gaussian system requires initial
data for the $1$-form $f$. For this one takes
\begin{equation}\label{initialf}
\langle f, \partial_\tau\rangle=0, \quad \pi^* f =\kappa^{-1}\mbox{d}\kappa. 
\end{equation}

The reduced field equations \eqref{redEq1}-\eqref{redEq3} and the
symmetric hyperbolic system implied by \eqref{eqConfFrame4} can be
interpreted as equations in the development of $\c{B}^*$. This
development will be denoted by $\tc{N}$, and is a $5$-dimensional
manifold smoothly embedded in the bundle of fame fields over
$\tc{M}$. The manifold $\tc{N}$ is a $SO(2)$ bundle over the
spacetime. Its projection sending $\tc{N}$ onto $\tc{M}$ will again be
denoted by $\pi$. The coordinates $\rho$, $s$ and the vector fields
$X$, $c_{\bm a}$ are pushed forward with the flow of the conformal geodesics
ruling $\tc{N}$, in such a way that $X$ generates the Kernel of
$\pi$. The parameter $x^0\equiv\tau$ defines a further independent
coordinate with $x^0=\tau=0$ on $\c{B}^*$. The tangent vector field of
this congruence is denoted by $\partial_\tau$. To interpret the
reduced field equations as equations on $\tc{N}$ it is assumed that
frame fields $e_{\bm\alpha}$ are vector fields on $\tc{N}$ which are defined at a
frame field $(\partial_\tau,c_{\bm a})$ by the requirements that: (i) they
project onto the frame field defined by $(\partial_\tau,c_{\bm a})$ on
$\tc{M}$; (ii) they do not pick up components in the $X$
direction. The unknowns in the reduced field equations are then
interpreted as vector-valued functions on $\tc{N}$.

\subsection{The conformal boundary}
From equations \eqref{confFact} and \eqref{initialf} it follows that
\begin{equation}\label{Thetasol}
 \Theta=\Theta_*\left(1-\tau^2\frac{\kappa_*^2}{\omega_*^2}\right),\hspace{1cm}\mbox{on}\,\,\tc{N},
\end{equation}
where the function $\omega$ is given by
\[
 \omega=\frac{2\Omega}{\sqrt{|D_a\Omega D^a\Omega|}},\hspace{1cm}\mbox{on}\,\,\c{B}^*.
\]
As before, subscripts `$*$' imply that the relevant functions are
constant along the conformal geodesics. Explicit expressions for
$d_{\bm\alpha}$ can be obtained from \eqref{compb} using that $f$ and
$\t{f}$ are related by $f=\t{f}-\Theta^{-1}\mbox{d}\Theta$.

\medskip
An important property of the construction described in the previous
lines is that if the initial data set for the reduced conformal field
equations has a smooth limit as $\rho\rightarrow 0$, then it can be
smoothly extended into the coordinate range $\rho\leq 0$. Similarly,
$\Theta$ and $d_{\bm\alpha}$ take smooth limits as $\rho\rightarrow 0$ and
can then be extended smoothly into a range where $\rho\leq0$. It
follows that the initial value problem for the reduced field equations
can be extended smoothly into a range where $\rho\leq0$ in such a way
that the reduced equations are still a symmetric hyperbolic system. If
the development of the initial value problem just formulated extends
far enough, then the following regions of the development can be
distinguished:
\begin{eqnarray*}
&& {\tc{N}}  \equiv  \left\{|\tau|<\frac{\omega}{\kappa},\,0<\rho<a,\,s\in SO(3)\right\},\\
&&\bar{\c{N}} \equiv  \left\{|\tau|\leq\frac{\omega}{\kappa},\,0\leq\rho<a,\,s\in SO(3)\right\},
\end{eqnarray*}
where $\omega/\kappa$ is a function of $\rho$ and $s$. One also
has the $4$-dimensional submanifolds
\begin{eqnarray*}
&&\scri^+  \equiv  \left\{\tau=+\frac{\omega}{\tau},\,0<\rho<a,\,s\in SO(3)\right\},\\
&& \scri^-  \equiv  \left\{\tau=-\frac{\omega}{\tau},\,0<\rho<a,\,s\in SO(3)\right\},\\
&& {\c{I}}  \equiv  \left\{\tau<1,\,\rho=0,\,s\in SO(3)\right\},
\end{eqnarray*}
and the $3$-dimensional submanifolds
\begin{eqnarray*}
&& {\c{I}}^+  \equiv  \left\{\tau=+1,\,\rho=0,\,s\in SO(3)\right\},\\
&& {\c{I}}^-  \equiv  \left\{\tau=-1,\,\rho=0,\,s\in SO(3)\right\},\\
&& {\c{I}}^0  \equiv \left\{\tau=0,\,\rho=0,\,s\in SO(3)\right\},
\end{eqnarray*}
where it has been observed that $\omega/\kappa\rightarrow 1$ as
$\rho\rightarrow 0$. It can be verified that 
\begin{eqnarray*}
&& \Theta>0\hspace{1cm}\mbox{on}\,\,\tc{N},\\
&& \Theta=0,\,\,\mbox{d}\Theta\neq0\hspace{1cm}\mbox{on}\,\,\scri^-\cup\scri^+\cup\c{I},\\
&& \Theta=0,\,\,\mbox{d}\Theta=0\hspace{1cm}\mbox{on}\,\,{\c{I}}^-\cup{\c{I}}^+.
\end{eqnarray*}
The initial hypersurface $\c{B}^*$ is given by
\[
 {\c{B}}^*=\left\{\tau=0,\,0<\rho<a,\,s\in SO(3)\right\}.
\]
Its closure in $\bar{\c{N}}$ is given by
\[
 \bar{\c{B}}\equiv\left\{\tau=0,\,0\leq\rho<a,\,s\in SO(3)\right\}={\c{B}}^*\cup{\c{I}}^0.
\]
Factoring out the group $SO(2)$ projects $\tc{N}$ onto the set
$\tc{M}$,  representing the ``physical spacetime''.

\medskip
\noindent
\textbf{Observation.} The data on $\c{B}^*$ have a unique smooth extension to
$\bar{\c{B}}$. Furthermore, the solution on $\bar{\c{N}}$ depends only on
this data as the set $\c{I}$ is a total characteristic of the
system of equations, and therefore the solution there depends only on
the data on ${\c{I}}^0$. The set $\c{I}$, referred to as the
\emph{cylinder at spacelike infinity}, represents a boundary of the
spacetime $\tc{N}$, and may be understood as a blow up of spacelike
infinity $i^0$. Of particular importance are the sets ${\c{I}}^+$ and
${\c{I}}^-$, \emph{the critical sets}, as the system of evolution equations
degenerates there. This degeneracy makes very difficult to make any
statement about smoothness of the solution to the initial value problem
stated above, even it has been shown that for general initial data
this sets are not regular ---see e.g. \cite{Fri98a,Fri04}.

\medskip
In view of the discussion of the previous paragraphs, the objective of
the present article is as follows: to show that for stationary
asymptotically flat initial data the solutions to the regular finite
initial value problem are smooth in a neighbourhood of $\c{I}$, and in
particular are smooth through ${\c{I}}^+$ and ${\c{I}}^-$.

\section{Stationary asymptotically flat spacetimes}\label{stationarySection}

In what follows, let $(\tc{M},\t{g}_{\mu\nu},\xi^\mu)$ denote a a
stationary spacetime. That is, $\tc{M}$ is a four-dimensional
manifold, $\t{g}_{\mu\nu}$ is a Lorentzian metric on $\tc{M}$ with signature
$(+,-,-,-)$ satisfying the Einstein vacuum field equations, and $\xi^\mu$
is a time-like Killing vector field with complete orbits. As discussed
in Section \ref{Subsection:IntroStationary}, instead of working with
the 4-dimensional manifold $\tc{M}$ it is more convenient to consider
the quotient manifold, $\tc{X}$, of $\tc{M}$ with respect to the
trajectories of the Killing vector $\xi^\mu$ ---see
\cite{BeiSch00,Ger71,Ger72b}. 

\subsection{The stationary metric in terms of quantities on the
  quotient manifold}
Locally, the metric $\t{g}_{\mu\nu}$ can be written
in terms of quantities defined on the quotient manifold $\tc{X}$: a scalar $V$, a 1-form
$\t{\beta}_a$, and a Riemannian metric $\t{\gamma}_{ab}$.  More precisely, one has that
\begin{equation}\label{metricStat}
 \t{g}_{\mu\nu}\mbox{d}\t{x}^\mu \mbox{d}\t{x}^\nu=V(\mbox{d}t+\t{\beta}_a\mbox{d}\t{x}^a)(\mbox{d}t+\t{\beta}_b\mbox{d}\t{x}^b)-V^{-1}\t{\gamma}_{ab}\mbox{d}\t{x}^a\mbox{d}\t{x}^b,
\end{equation}
where $V$, $\t{\beta}_a$ and $\t{\gamma}_{ab}$ depend only on the spatial coordinates $\t{x}^a$.

\medskip
In order to obtain the field equations implied on $\tc{X}$ by Einstein
vacuum field equations on $\tc{M}$ it is convenient to consider the
quantity $\t{\omega}_a$, defined on $\tc{X}$ by
\begin{equation}\label{om_i}
 \t{\omega}_a=-V^2\t{\epsilon}_{abc}\t{D}^b\t{\beta}^c,
\end{equation}
where $\t{D}$ is the covariant derivative associated with
$\t{\gamma}_{ab}$. Then the Einstein vacuum field equations on $\tc{M}$ imply
\[
 \t{D}_{[a}\t{\omega}_{b]}=0.
\]
If one further assumes $\tc{X}$ to be simply connected (our analysis
will concentrate in a neighbourhood of infinity), then there exists a
scalar field $\omega$ such that
\[
 \t{D}_a\omega=\t{\omega}_a.
\]
In the sequel we will consider `gauge' transformations of the form
\[
 \t{\beta}_a\rightarrow\t{\beta}_a+\partial_af,
\]
where $f$ is a scalar field on $\tc{X}$. Clearly, $\tc{\omega}_a$ does
not change under these transformations ---cfr. equation \eqref{om_i}.
Moreover, the metric remains unchanged if one sets  $t\rightarrow t-f$.

\subsection{The Hansen potentials}
In order to write down the stationary field equations it is convenient to
introduce the so-called \emph{Hansen potentials}:
\begin{equation*}
 \t{\phi}_M =  \frac{V^2+\omega^2-1}{4V},\hspace{1cm}\t{\phi}_S  =  \frac{\omega}{2V},\hspace{1cm}\t{\phi}_K  =  \frac{V^2+\omega^2+1}{4V}.
\end{equation*}
They are not independent as 
\[
 \t{\phi}_M^2+\t{\phi}_S^2-\t{\phi}_K^2=-\tfrac{1}{4}.
\]
The vacuum field equations on $\tc{M}$ then imply on $\tc{X}$
\begin{subequations}
\begin{eqnarray}
&&\t{\Delta}\t{\phi}_\alpha= 
2\t{R}[\t{\gamma}]\t{\phi}_\alpha,\hspace{1cm}\alpha=M,S,K, \label{fieldEqPhi}\\
&&\t{R}_{ab}[\t{\gamma}] = 
2(\t{D}_a\t{\phi}_M\t{D}_b\t{\phi}_M+\t{D}_a\t{\phi}_S\t{D}_b\t{\phi}_S-\t{D}_a\t{\phi}_K\t{D}_b\t{\phi}_K). \label{fieldEqR}
\end{eqnarray}
\end{subequations}
The latter will be regarded as field equations for $\t{\gamma}_{ab}$,
$\t{\phi}_M$ and $\t{\phi}_S$ on $\tc{X}$. They are equivalent to
Einstein vacuum field equations on $\tc{M}$ in the sense that $\tc{M}$
can be reconstructed as a stationary spacetime if
$\t{\gamma}_{ab}$, $\t{\phi}_M$ and $\t{\phi}_S$ are given.

\subsection{The 3+1 form of the stationary metric}
As mentioned in Section \ref{Subsection:IntroStationary} of the
introduction, although the field equations take the simple
form \eqref{fieldEqPhi}-\eqref{fieldEqR} in $\tc{X}$, our main
interest is to consider the Cauchy problem with data arising from
stationary spacetimes. For this one needs to consider the $3+1$
decomposition of the spacetime metric $\tilde{g}_{\mu\nu}$ with
respect to a particular spacelike hypersurface of $\tc{M}$. If we
choose $\tc{S}$ as defined by $t=\mbox{constant}$, then $\t{g}_{\mu\nu}$ has a
$3+1$ decomposition with respect to $\tc{S}$ given by
\begin{equation}
\label{metricDec} \t{g}_{\mu\nu}\mbox{d}\t{x}^\mu
\mbox{d}\t{x}^\nu=\t{N}^2\mbox{d}t^2-\t{h}_{ab}(\t{N}^a\mbox{d}t+\mbox{d}\t{x}^a)(\t{N}^b\mbox{d}t+\mbox{d}\t{x}^b),
\end{equation} 
where $\t{N}$, $\t{N}^a$, $\t{h}_{ab}$ denote,
respectively, the lapse function, the shift vector and the intrinsic
metric of the hypersurface $\tc{S}$. Comparing the line
elements \eqref{metricStat} and \eqref{metricDec} one finds that
\begin{equation}\label{Translation}
 \t{N}^2 =
\frac{V}{1-V^2\t{\beta}_a\t{\beta}^a},\hspace{1cm}\t{N}^a =
-\frac{V^2\t{\beta}^a}{1-V^2\t{\beta}_b\t{\beta}^b},\hspace{1cm}\t{h}_{ab} = V^{-1}\t{\gamma}_{ab}-V\t{\beta}_a\t{\beta}_b.
\end{equation}
We will adopt the convention of moving indices of
objects in $\tc{X}$ with the quotient metric $\t{\gamma}_{ab}$. The
relations \eqref{Translation}
allow us to go
back and forth between quantities defined on $\tc{X}$ and quantities
defined on $\tc{S}$.

\subsection{Asymptotic flatness and its consequences}
Following the usual assumptions about asymptotic flatness for
stationary spacetimes, it will be assumed that
$(\tc{X},\t{\gamma}_{ab},\t{\phi}_M,\t{\phi}_S)$ is asymptotically
Euclidean and regular in the sense discussed in Section
\ref{Subsection:AsymptoticallyEuclidean} of the introduction ---see
e.g. \cite{BeiSim81a,BeiSim81b}. More precisely, it is assumed that
there exists a manifold $\c{X}$, such that ${\c{X}}=\tc{X}\cup\{i\}$,
where $i$ is a point. Furthermore, it is assumed that 
for some real constant $B^2>0$ the conformal factor
\begin{equation}\label{defOm}
 \Omega=\tfrac{1}{2}B^{-2}[(1+4\t{\phi}_M^2+4\t{\phi}_S^2)^{\tfrac{1}{2}}-1]
\end{equation}
is $C^{2,\alpha}$ on $\c{X}$ and satisfies
\[
 \Omega(i)=0,\hspace{1cm}D_a\Omega(i)=0.
\]
In addition it will be assumed that
\begin{equation}
\gamma_{ab}=\Omega^2\t{\gamma}_{ab} \label{PhysicalQuotientToConformalQuotient}
\end{equation}
extends to a $C^{4,\alpha}$
metric on $\c{X}$ and satisfies 
\[
 D_aD_b\Omega(i)=2\gamma_{ab}(i),
\]
where $D$ is the Levi-Civita covariant derivative of the 3-metric
$\gamma_{ab}$. 

\medskip
The conformal rescaling of the metric given by \eqref{PhysicalQuotientToConformalQuotient}
suggests the following definition of rescaled potentials:
\[
 \phi_\alpha=\Omega^{-\tfrac{1}{2}}\t{\phi}_\alpha,\hspace{2cm}\alpha=M,S,K.
\]
The motivation behind the introduction of conformally rescaled fields
is the following theorem by Beig \& Simon \cite{BeiSim81a} ---see also
\cite{Kun81a}.

\begin{theorem}[Theorem 1 of \cite{BeiSim81a}]
\label{Theorem:BeigSimon}
 For any asymptotically flat solution
$(\t{\gamma}_{ab},\t{\phi}_M,\t{\phi}_S)$ of the stationary equations
\eqref{fieldEqPhi}-\eqref{fieldEqR} there exists a chart defined in
some neighbourhood of $i$ in $\c{X}$ such that
$(\gamma_{ab},\phi_M,\phi_S,\Omega)$ are analytic.
\end{theorem}

\medskip
\noindent
\textbf{Remark.} Given the chart indicated by the previous theorem,
one can make a coordinate transformation to $\gamma$-normal
coordinates $x^a$ centered at $i$. The fields
$(\gamma_{ab},\phi_M,\phi_S,\Omega)$ are also analytic with respect to
the normal coordinates $x^a$ ---this follows from the
Cauchy-Kovalewska theorem applied to the equations of the radial
geodesics written in the analytic coordinates given by Theorem \ref{Theorem:BeigSimon}. It is important to notice that
Theorem \ref{Theorem:BeigSimon}  does not make any assertion about the smoothness of other
quantities on $\cal{X}$, like $V$, $\phi_K$, $\beta_a$ or quantities
defined on a hypersurface of the spacetime. An analysis of the
regularity of these and other related quantities has been carried out
in \cite{Dai01b}. 

\medskip
In the sequel, we will require several results from
\cite{Dai01b}. These will be presented here for completeness and quick
reference. In the following let the radial coordinate $\rho$ be
defined by
\begin{equation}
\rho\equiv\left(\sum^3_{i=1}(x^a)^2\right)^{1/2}.
\label{Definition:rho}
\end{equation}

 In \cite{Dai01b} it was found that the non-analyticity of the
 relevant functions is of a very special type and it depends only on the
 coordinate $\rho$. Accordingly, one defines the following function space:

\begin{definition}[Definition 2.2 of \cite{Dai01b}]
 We define the space $E^{\omega}$ as the set 
\[
E^{\omega}=\{f=f_1+\rho f_2:f_1,f_2\in C^\omega\},
\]
where $C^\omega$ denotes the set of analytic functions in a
neighbourhood of $i$.
\end{definition}
Associated to the latter definition one has the following:

\begin{lemma}[Lemma 2.3 of \cite{Dai01b}]
Let $f,g\in E^{\omega}$, then
\begin{itemize}
 \item[(i)] $f+g\in E^{\omega}$.
 \item[(ii)] $fg\in E^{\omega}$.
 \item[(iii)] If $f\neq 0$ then $1/f\in E^{\omega}$.
\end{itemize}
\end{lemma}

Obviously, if $f\in C^{\omega}$ then
$f\in E^{\omega}$. The main result of the analysis in \cite{Dai01b} is
that most of the relevant quantities belong to $E^{\omega}$. In
particular, one has that:

\begin{lemma}[Lemmas 2.4 and 2.5 of \cite{Dai01b}]
\label{expBeta}
In the normal coordinates implied by Theorem \ref{Theorem:BeigSimon}
one has that
\[
V\in E^{\omega}.
\]
Furthermore, there exist a choice of gauge for which the 1-form
$\beta_a$ has the following form:
\begin{equation}\label{beta}
 \beta_a=\beta^1_a+\frac{\beta^2_a}{\rho},
\end{equation}
where $\beta^1_a$, $\beta^2_a$ are analytic functions of $x^a$ given by
\begin{equation}\label{beta12}
 \beta^1_a=e_{abc}\,f_1^b\,x^c,\hspace{2cm}\beta^2_a=e_{abc}\,f_2^b\,x^c,
\end{equation}
where $f_1^a$, $f_2^a$ are analytic and $e_{abc}$ is the flat volume
element. In particular, it follows that $\beta_ax^a=0$.
\end{lemma}

\subsection{A first conformal compactification of stationary spacetimes}
As discussed in \cite{Dai01b}, the fields $\Omega$ and $V$ can be used
to construct a first conformal compactification of the \emph{physical manifolds}
$\tc{M}$ and $\tc{S}$. For this one introduces a conformal factor $\breve{\Omega}$ via
\[
 \breve{\Omega}\equiv V^{1/2}\Omega.
\]
Note that, as defined,  $\breve{\Omega}$ is not analytic as $V$ is not
analytic. The associated rescaled metrics are then given by
\[
 \breve{g}_{\mu\nu}=\breve{\Omega}^2\t{g}_{\mu\nu},\quad \breve{h}_{ab}=\breve{\Omega}^2\t{h}_{ab}.
\]
Hence, one has that
\begin{equation}\label{stat_metric_gamma}
 \breve{g}_{\mu\nu}\mbox{d}x^\mu \mbox{d}x^\nu=V^2\Omega^2(\mbox{d}t+\beta_a\mbox{d}x^a)(\mbox{d}t+\beta_b\mbox{d}x^b)-\gamma_{ab}\mbox{d}x^a\mbox{d}x^b.
\end{equation}
Moreover, one also has a $3+1$ decomposition with respect to the hypersurfaces $\tc{S}=\{t=\mbox{constant}\}$,
\begin{equation}\label{stat_metric_h}
 \breve{g}_{\mu\nu}\mbox{d}x^\mu \mbox{d}x^\nu=N^2\mbox{d}t^2-\breve{h}_{ab}(N^a\mbox{d}t+\mbox{d}x^a)(N^b\mbox{d}t+\mbox{d}x^b).
\end{equation}
By comparison with equation \eqref{metricDec} one gets that
\begin{subequations}
\begin{eqnarray}
&& N=\breve{\Omega}\t{N},\hspace{1cm}N^a=\t{N}^a, \label{LapseShift}\\
&& \breve{h}_{ab}=\gamma_{ab}-V^2\Omega^2\beta_a\beta_b, \label{3-metric}
\end{eqnarray}
\end{subequations}
where $\breve{h}_{ab}$ is the intrinsic metric on $\tc{S}$. Fundamental
for our subsequent analysis is the following result. 

\begin{theorem}[Theorem 2.6 of \cite{Dai01b}]
 Assume $\beta_a$ is given by Lemma \ref{expBeta}. Then, in some neighbourhood of $i$, the metric $\breve{h}_{ab}$ has the form
\begin{equation}\label{metric}
 \breve{h}_{ab}=\breve{h}^1_{ab}+\rho^3\breve{h}^2_{ab},
\end{equation}
where $\breve{h}^1_{ab},\;\breve{h}^1_{ab}\in C^\omega$.
\end{theorem}

\medskip
\noindent
\textbf{Remark.} The later result implies that the conformal 3-metric $\breve{h}_{ab}$ is in
$C^{2,\alpha}$. Therefore, the conformal factor $\breve{\Omega}$ can be
used to define a conformal compactification $\c{S}$ of the Cauchy
slice $\tc{S}$ plus the point at infinity $i$, in the same way as made
for $\tc{X}$. This implies that the pair $(\tc{S},\t{h}_{ab})$ admits a
$C^{2,\alpha}$ compactification ---that is, the pair is asymptotically Euclidean and
regular in the sense discussed in Section \ref{Subsection:AsymptoticallyEuclidean} of the Introduction. The decomposition of
$\beta_a$ given by equation \eqref{beta}  in Lemma \ref{expBeta}, and
thus, the decomposition of $\breve{h}$ given by equation \eqref{metric} is
preserved under the transformation
\[
 \beta_a\rightarrow\beta_a+\partial_af,\hspace{1cm}f\in E^\omega.
\]
If one imposes the condition $x^a\beta_a=0$, then one fixes
$\partial_af$. Accordingly, $\beta_a$, as given by equation \eqref{beta} is the unique
possible choice of the 1-form $\beta_a$ that satisfies this condition.

\medskip
Let $\t{\chi}_{ab}$, $\breve{\chi}_{ab}$ denote, respectively, the extrinsic curvatures of $\tc{S}$ with
respect to the metrics $\t{g}_{\mu\nu}$ and $\breve{g}_{\mu\nu}$. One has that
$\t{\chi}_{ab}=\breve{\Omega}^{-1}\breve{\chi}_{ab}$. Furthermore, we define 
\[
 \breve{\psi}_{ab}\equiv \breve{\Omega}^{-1}\t{\chi}_{ab}=\breve{\Omega}^{-2}\breve{\chi}_{ab}.
\]
The behaviour of $\breve{\psi}_{ab}$ near $i$ is given by the following result.
\begin{theorem}[Theorem 2.7 of \cite{Dai01b}]
 Assume $\beta_a$ as given by Lemma \ref{expBeta}. Then in some neighbourhood of $i$, the tensor $\breve{\psi}_{ab}$ has the form
\[
 \breve{\psi}_{ab}=\rho^{-5}f\,x_{(a}\beta^2_{b)}+\rho^{-3}\breve{\psi}^1_{ab},
\]
where $\breve{\psi}^1_{ab}\in E^\omega$, $f\in E^\omega$ and $\beta_a^2$ is given by \eqref{beta12}. Furthermore, $\rho^8\breve{\psi}_{ab}\breve{\psi}^{ab}\in E^\omega$.
\end{theorem}

\subsection{Detailed expansions at infinity}

Key for our present analysis is that if a quantity belongs to the
space $E^{\omega}$, then although it is not analytic, it nevertheless
has in a neighbourhood of infinity an analytic expansion in terms of
the radial coordinate $\rho$ and the angular coordinates. In the
sequel it will be necessary not only to know that relevant fields
belong to $E^{\omega}$, but also to know the first orders of the
expansion in a neighbourhood of $i$. In \cite{SimBei83} the first
orders of the asymptotic expansions of the unrescaled fields $\t{\phi}_M$, $\t{\phi}_S$, $\t{\phi}_K$
and $\t{\gamma}_{ab}$ has been explicitly given in terms of constant tensors $M$, $S$, $M_a$, $S_a$, $M_{ab}$, $S_{ab}$, etc. These expansions read
\begin{eqnarray*}
&& \t{\phi}_M =  \frac{M}{\t{r}}+\frac{M_a\t{x}^a}{\t{r}^3}+\frac{M_{ab}\t{x}^a\t{x}^b}{2\t{r}^5}+\frac{M(M^2+S^2)}{\t{r}^3}+O(\t{r}^{-4}),\\
&& \t{\phi}_S =  \frac{S}{\t{r}}+\frac{S_a\t{x}^a}{\t{r}^3}+\frac{S_{ab}\t{x}^a\t{x}^b}{2\t{r}^5}+\frac{S(M^2+S^2)}{\t{r}^3}+O(\t{r}^{-4}),\\
&& \t{\phi}_K =  \frac{1}{2}+\frac{M^2+S^2}{\t{r}^2}+\frac{2MM_ax^a}{\t{r}^4}+\frac{2SS_ax^a}{\t{r}^4}+O(\t{r}^{-4}),\\
&& \t{\gamma}_{ab}  =  \delta_{ab}-\frac{M^2}{\t{r}^4}(\delta_{ab}\t{r}^2-\t{x}_a\t{x}_b)-\frac{2MM_{(a}\t{x}_{b)}}{\t{r}^4}-\frac{2MM_c\t{x}^c\delta_{ab}}{\t{r}^4}+\frac{4MM_c\t{x}^c\t{x}_a\t{x}_b}{\t{r}^6}\\
&&\hspace{1.5cm}+\frac{S^2}{\t{r}^4}(\delta_{ab}\t{r}^2+\t{x}_a\t{x}_b)+\frac{2SS_{(a}\t{x}_{b)}}{\t{r}^4}+\frac{2SS_c\t{x}^c\delta_{ab}}{\t{r}^4}-\frac{4SS_c\t{x}^c\t{x}_a\t{x}_b}{\t{r}^6}+O(\t{r}^{-4}),
\end{eqnarray*}
where indices in the coordinates and constant tensors are moved with
the flat metric $\delta_{ab}$. 

\medskip
\noindent
\textbf{Remark}. Asymptotic flatness implies that the angular momentum
monopole $S$ has to vanish. Furthermore, by a suitable choice of the
origin for the coordinates $\tilde{x}^a$ one can make
$M_a=0$. In order to simplify our computations we assume this choice
of coordinates.

\medskip
Following the discussion in the previous subsections, we want to
consider the fields in a neighbourhood of $i$ written in terms of normal coordinates
centered at $i$. This is done in several steps. First, one changes 
coordinates from $\tilde{x}^a$ to $y^a$, via
\[
 \tilde{x}^a=-\frac{y^a}{r^2},\quad r^2=\delta_{ab}y^ay^b=\frac{1}{\t{r}^2}.
\]
Then, using the expression \eqref{defOm} for the conformal factor with
$B=M$, one finds that
\[
 \Omega=r^2+M^2r^4+\frac{1}{M^2}(S_ay^a)^2r^2+\frac{1}{M}M_{ab}y^ay^br^2+O(r^5).
\]

\medskip
Furthermore, from the definition of the conformal potentials if
follows that
\begin{eqnarray*}
&& \phi_M =  M+\frac{1}{2}M^3r^2-\frac{1}{2M}(S_ay^a)^2+O(r^3),\\
&& \phi_S =  -S_ay^a+\frac{1}{2}S_{ab}y^ay^b+O(r^3),\\
&& \phi_K  =  \frac{1}{2r}+\frac{3}{4}M^2r-\frac{1}{4M^2r}(S_ay^a)^2-\frac{1}{4Mr}M_{ab}y^ay^b+O(r^2).
\end{eqnarray*}
Next, we give the expansion for the conformally rescaled metric
$\gamma_{ab}$. However, instead of giving these expansions with
respect to the coordinates $y^a$, we use normal coordinates, $x^a$, centred at
$i$ ---as it was done in the remark after Theorem
\ref{Theorem:BeigSimon}.  For this one requires that after the coordinate transformation $y^a\rightarrow x^a$ the metric satisfies
\[
\gamma_{ab}\,x^b=\delta_{ab}\,x^b.
\]
A lengthy computation renders
\begin{eqnarray*}
&&\gamma_{ab}  =  \delta_{ab}+\frac{\rho^2}{3M^2}\Big(M^4(\delta_{ab}-e_ae_b)+2M\big(M_{ab}+\delta_{ab}M_{cd}e^ce^d-2e_{(a}M_{b)c}e^c\big)\\
&&\hspace{2.5cm}+2\big(S_aS_b+\delta_{ab}(S_ce^c)^2-2e_{(a}S_{b)}S_ce^c\big)\Big)+O(\rho^3),
\end{eqnarray*}
where $\rho$ is defined by \eqref{Definition:rho} and $e^a\equiv x^a/\rho$. 
The leading terms of the expansion of the inverse metric are given by
\begin{eqnarray*}
&&\gamma^{ab} =  \delta^{ab}-\frac{\rho^2}{3M^2}\Big(M^4(\delta^{ab}-e^ae^b)+2M\big(M^{ab}+\delta^{ab}M_{cd}e^ce^d-2e^{(a}M^{b)c}e_c\big)\\
&&\hspace{2.5cm}+2\big(S^aS^b+\delta^{ab}(S_ce^c)^2-2e^{(a}S^{b)}S_ce^c\big)\Big)+O(\rho^3),
\end{eqnarray*}
The transformation between the coordinates $y^a$ and $x^a$ is given
by 
\begin{subequations}
\begin{eqnarray}
&&y^a  =
x^a+\frac{1}{3M^2}\rho^3\Big(-M^4e^a+M\big(M^{ab}e_b-2e^aM_{bc}e^be^c\big)
\nonumber\\
&&\hspace{2.5cm} +S^aS_be^b-2e^a(S_be^b)^2\Big)+O(\rho^4), \label{CoordTransf1}\\
&&
r=\rho-\frac{1}{3M^2}\rho^3\Big(M^4+MM_{ab}e^ae^b+(S_ae^a)^2\Big)+O(\rho^4). \label{CoordTransf2}
\end{eqnarray}
\end{subequations}
Using \eqref{CoordTransf1}-\eqref{CoordTransf2}, one can express the
rescaled potentials in normal coordinates. One finds that
\begin{eqnarray*}
&&\phi_M  =  M+\frac{1}{2M}\rho^2(M^4-(S_ae^a)^2)+O(\rho^3),\\
&&\phi_S  =  -S_ae^a\rho+\frac{1}{2}\rho^2S_{ab}e^ae^b+O(\rho^3),\\
&&\phi_K =  \frac{1}{2\rho}+\frac{1}{12M^2}\rho(11M^4-MM_{ab}e^ae^b-(S_ae^a)^2)+O(\rho^2).
\end{eqnarray*}
For later use it is also convenient to calculate the expansion of
other quantities. Namely,
\begin{eqnarray*}
&& V =  1+2M\rho+2M^2\rho^2+\frac{1}{3M}\rho^3(4M^4+MM_{ab}e^ae^b-2(S_ae^a)^2)+O(\rho^4), \\
&& \beta_a=e_{abc}\,e^c (2S^b+O(\rho)).
\end{eqnarray*}
Using the formula \eqref{stat_metric_gamma} one finds that the
4-dimensional spacetime metric and its inverse are given by
\begin{eqnarray*}
&& (\breve{g}_{\mu\nu})=\left(\begin{array}{cc} V^2\Omega^2 &
    V^2\Omega^2\beta_a \\ V^2\Omega^2\beta_b &
    -\gamma_{ab}+V^2\Omega^2\beta_a\beta_b \end{array}\right), \\
&&(\breve{g}^{\mu\nu})=\left(\begin{array}{cc} \displaystyle\frac{1-V^2\Omega^2\beta_c\beta^c}{V^2\Omega^2} & \beta^a \\ \beta^b & -\gamma^{ab} \end{array}\right),
\end{eqnarray*}
where $\beta^a\equiv \gamma^{ab}\beta_b$. Moreover, one has the
following expansions for the components of the metric and its inverse:
\begin{eqnarray*}
&&\breve{g}_{tt} =  \rho^4+4M\rho^5+\frac{2}{3M^2}\rho^6\big(13M^4+MM_{ab}e^ae^b+(S_ae^a)^2\big)+O(\rho^7),\\
&& \breve{g}_{ta} =  \rho^4e_{abc}e^c (2S^b+O(\rho)),\\
&& \breve{g}_{ab} =  -\delta_{ab}-\frac{\rho^2}{3M^2}\Big(M^4(\delta_{ab}-e_ae_b)+2M\big(M_{ab}+\delta_{ab}M_{cd}e^ce^d-2e_{(a}M_{b)c}e^c\big)\\
&& \hspace{2.5cm}+2\big(S_aS_b+\delta_{ab}(S_ce^c)^2-2e_{(a}S_{b)}S_ce^c\big)\Big)+O(\rho^3),\\
&& \breve{g}^{tt} =  \frac{1}{\rho^4}-\frac{4M}{\rho^3}+\frac{2}{3M^2\rho^2}\big(11M^4-MM_{ab}e^ae^b-(S_ae^a)^2\big)+O(\rho^{-1}),\\
&& \breve{g}^{ta} =  e^{abc}e_c (2S_b+O(\rho)),\\
&& \breve{g}^{ab}  =  -\delta^{ab}+\frac{\rho^2}{3M^2}\Big(M^4(\delta^{ab}-e^ae^b)+2M\big(M^{ab}+\delta^{ab}M_{cd}e^ce^d-2e^{(a}M^{b)c}e_c\big)\\
&& \hspace{2.5cm}+2\big(S^aS^b+\delta^{ab}(S_ce^c)^2-2e^{(a}S^{b)}S_ce^c\big)\Big)+O(\rho^3).
\end{eqnarray*}
One also finds that
\begin{eqnarray*}
 &&\breve{\Omega} =  V^{1/2}\Omega =  \rho^2+M\rho^3+\frac{1}{3M^2}\rho^4\Bigg(\frac{5}{2}M^4+MM_{ab}e^ae^b+(S_ae^a)^2\Bigg)+O(\rho^5).
\end{eqnarray*}

\section{Conformal extension of stationary vacuum spacetimes}
\label{confExtSVS}

In this section we introduce a conformal extension of vacuum
stationary spacetimes which is well adapted for the analysis of the
structure of spatial infinity. To this end, we start from the
conformal extension of the stationary metric given by the line element
\eqref{stat_metric_gamma}. In order to analyse the regularity of the relevant fields we will make use of the first orders expansions in a neighbourhood of $i$, which we collect as we consider the corresponding quantities.  The present analysis is based on a similar
analysis for static spacetimes given in \cite{Fri04}. 

\medskip
Recall that $x^a$, $a=1,2,3$, are normal coordinates of the quotient
metric $\gamma_{ab}$. We have defined
\begin{equation*}
\rho\equiv \left(\sum_{a=1}^3(x^a)^2\right)^{1/2},\hspace{2cm}e^a\equiv \frac{x^a}{\rho} \quad \mbox{  for  }\rho>0.
\end{equation*}
For constant $t$, the surfaces of constant $\rho$ are diffeomorphic to
a 2-dimensional sphere. Accordingly,  arbitrary coordinates $\psi^A$, $A=2,3$ on
the 2-sphere $\Sphere^2=\{|x|=1\}$ can be used to parametrize
$e^a$. We
then write 
\[
e^a=e^a(\psi^A), \quad \mbox{d}e^a=e^a{}_{,\psi^A}\,\mbox{d}\psi^A.
\]
The coordinates $\psi^A$ can be chosen such that $e^a$
depends analytically on them. Consistent with the previous definitions one has that  $x^a=\rho\,
e^a(\psi^A)$. Therefore, the conformal 3-metric $\breve{h}_{ab}$ given
by equation \eqref{3-metric} takes the form
\[
 \breve{h}=\mbox{d}\rho^2+\rho^2k,
\]
where $k$ denotes the 2-dimensional metric on the surfaces of constant
$\rho$ given by 
\[
 k=k_{AB}\mbox{d}\psi^A\mbox{d}\psi^B\equiv \breve{h}_{ab}(\rho)\mbox{d}e^a\mbox{d}e^b.
\]

\medskip
\noindent
\textbf{Remark.} Notice that as $\rho\rightarrow 0$, the metric
$\breve{h}$ approaches the standard Euclidean metric in normal
coordinates and the metric $k$ approaches the standard line element
$\mbox{d}\sigma^2=k(0,\psi^A)$ on the $2$-dimensional unit sphere in
the coordinates $\psi^A$.

\subsection{Coordinates for the analysis of the cylinder at spatial infinity}

The coordinates $(t,\rho,\psi^A)$ are well adapted to the description
of spatial infinity as a point. In order to resolve the
structure of the cylinder at spatial infinity, it is convenient to
introduce new coordinates $(\bar{\tau},\bar{\rho},\psi^A)$. The
coordinate change is inspired by an analysis of the Minkowski
spacetime ---see e.g. \cite{Fri98a,Val03a}.

\medskip
We define $x^0=t$, $x^{0'}=\bar{\tau}$, $x^{1'}=\bar{\rho}$, $x^{A'}=\psi^A$ and consider the map $\Phi:x^{\mu'}\rightarrow x^\mu(x^{\mu'})$ given by
\begin{eqnarray*}
&& t(x^{\mu'})=x^0(x^{\mu'})=\int_{(1-\bar{\tau})\bar{\rho}}^{\bar{\rho}}\frac{\mbox{d}s}{(V\Omega)[se^a(\psi^A)]},\\
&& x^a(x^{\mu'})=(1-\bar{\tau})\bar{\rho}e^a[\psi^A],
\end{eqnarray*}
where the squared brackets indicate the arguments of a given
function.  One explicitly finds that
\begin{subequations}
\begin{eqnarray}
 &&\mbox{d}t=\frac{\bar{\rho}}{(V\Omega)[(1-\bar{\tau})\bar{\rho}e^a]}\mbox{d}\bar{\tau}+\Bigg(\frac{1}{(V\Omega)[\bar{\rho}e^a]}-\frac{1-\bar{\tau}}{(V\Omega)[(1-\bar{\tau})\bar{\rho}e^a]}\Bigg)\mbox{d}\bar{\rho}+l, \label{dt}\\
&& 
 \mbox{d}x^a=-\bar{\rho}e^a\mbox{d}\bar{\tau}+(1-\bar{\tau})e^a\mbox{d}\bar{\rho}+(1-\bar{\tau})\bar{\rho}\,\mbox{d}e^a, \label{dxa}
\end{eqnarray}
\end{subequations}
where
\[
 l=l_A\mbox{d}\psi^A,\hspace{1cm}l_A=\int_{(1-\bar{\tau})\bar{\rho}}^{\bar{\rho}}\Bigg(\frac{1}{(V\Omega)[se^a]}\Bigg)_{,\psi^A}\mbox{d}s,\hspace{1cm}\mbox{d}e^a=(e^a)_{,\psi^A}\mbox{d}\psi^A.
\]
The differentials $\mbox{d}t$, $\mbox{d}x^a$ are independent for
$0\leq\bar{\tau}<1$ and $\bar{\rho}$ between zero and a small enough
number. Therefore one can consider the $x^{\mu'}$ as smooth
coordinates on an open neighbourhood of space-like infinity in
$\{t\geq0\}$. For later use we notice the relation
\[
 \rho=(1-\bar{\tau})\bar{\rho},
\]
which will be used to simplify the notation.

\medskip
The main purpose of the coordinate transformation is to remove the
coefficient $V^2\Omega^2$ from the time-time component of the metric
$\breve{g}_{\mu\nu}$ and to introduce a convenient parametrisation set
at which $\rho=0$, so that it seems to have an extension in the time direction.

\subsection{A frame adapted to spatial infinity}
\label{Subsection:Frame}

We define a set of frame fields $v_{\bm\alpha}$ and their associated coframe fields $\alpha^{\bm\alpha}$ by
\begin{eqnarray*}
&& v_{\bm 0}=\partial_{\bar{\tau}}, \quad
v_{\bm 1}=\bar{\rho}\partial_{\bar{\rho}}, \quad v_{\bm A}=\partial_{\psi^A}, \\
&& \alpha^{\bm 0}=\mbox{d}\bar{\tau},
\quad\alpha^{\bm1}=\frac{1}{\bar{\rho}}\mbox{d}\bar{\rho},\quad
\alpha^{\bm A}=\mbox{d}\psi^A.
\end{eqnarray*}
To change the frame field associated to our original coordinates to
the frame $v_{\bm\alpha}$ one needs the inner products
$v^\mu{}_{\bm \alpha}=\langle \mbox{d}x^\mu,v_{\bm\alpha}\rangle$. From the expressions \eqref{dt} and \eqref{dxa} one sees that:
\begin{eqnarray*}
&& v^t{}_{\bm 0} = \frac{\bar{\rho}}{(V\Omega)[\rho e^a]} \\
&& \phantom{v^t{}_0}=\frac{\bar{\rho}}{\rho^2}-2M\frac{\bar{\rho}}{\rho}+\frac{1}{3M^2}\bar{\rho}\big(5M^4-MM_{ab}e^ae^b-(S_ae^a)^2\big)+O(\bar{\rho}^2),\\
&& v^t{}_{\bm 1} =  \frac{\bar{\rho}}{(V\Omega)[\bar{\rho}e^a]}-\frac{\rho}{(V\Omega)[\rho e^a]} \\
&& \phantom{v^t\,_1 }=  -\frac{\bar{\tau}}{\rho}+\frac{1}{3M^2}\bar{\tau}\bar{\rho}\big(5M^4-MM_{ab}e^ae^b-(S_ae^a)^2\big)+O(\bar{\rho}^2),\\
&&v^t{}_{\bm A} =  l_A, \\
&&v^a{}_{\bm 0} =  -\bar{\rho}\,e^a, \\
&&v^a{}_{\bm 1} =  \rho\,e^a, \\
&&v^a{}_{\bm A} =  \rho\,e^a{}_{,\psi^A}. \\
\end{eqnarray*}

\medskip
The frame and coframe fields distort the length of the radial
component of the tensorial fields they are contracted with. This
distortion will be of importance in the sequel when discussing 
objects that are singular at spatial infinity.

\subsection{A conformal metric containing the cylinder at spatial infinity}

Let  $\bar{g}_{\mu\nu}$ denote a metric conformal to $\breve{g}_{\mu\nu}$ and $\t{g}_{\mu\nu}$ defined by
\[
 \bar{g}_{\mu\nu}=\frac{1}{\rho^2}\breve{g}_{\mu\nu}=\bar{\Omega}^2\t{g}_{\mu\nu}.
\]
It follows that 
\[
 (\bar{g}_{\mu\nu})=\left(\begin{array}{cc} V^2\Omega^2/\rho^2 & V^2\Omega^2\beta_a/\rho^2 \\ V^2\Omega^2\beta_b/\rho^2 & (-\gamma_{ab}+V^2\Omega^2\beta_a\beta_b)/\rho^2 \end{array}\right),
\]
and
\begin{eqnarray*}
&& \bar{g}_{tt} =  \rho^2+4M\rho^3+\frac{2}{3M^2}\rho^4\big(13M^4+MM_{ab}e^ae^b+(S_ae^a)^2\big)+O(\rho^5),\\
&& \bar{g}_{ta} =  \rho^2e_{abc}(2S^b+O(\rho))e^c,\\
&& \bar{g}_{ab} =  -\frac{1}{\rho^2}\delta_{ab}-\frac{1}{3M^2}\Big[M^4(\delta_{ab}-e_ae_b)+2M\big(M_{ab}+\delta_{ab}M_{cd}e^ce^d-2e_{(a}M_{b)c}e^c\big)\\
&& \hspace{2cm} +2\big(S_aS_b+\delta_{ab}(S_ce^c)^2-2e_{(a}S_{b)}S_ce^c\big)\Big]+O(\rho).
\end{eqnarray*}
The inverse metric $\bar{g}^{\mu\nu}$ is given by
\[
 (\bar{g}^{\mu\nu})=\left(\begin{array}{cc}
     \rho^2(1-V^2\Omega^2\beta_c\beta^c)/V^2\Omega^2 & \rho^2\beta^a
     \\ 
\rho^2\beta^b & -\rho^2\gamma^{ab} \end{array}\right),
\]
and
\begin{eqnarray*}
&& \bar{g}^{tt} =  \frac{1}{\rho^2}-\frac{4M}{\rho}+\frac{2}{3M^2}\big(11M^4-MM_{ab}e^ae^b-(S_ae^a)^2\big)+O(\rho^0),\\
&& \bar{g}^{ta} =  \rho^2e^{abc}(2S_b+O(\rho))e_c,\\
&& \bar{g}^{ab} =  -\rho^2\delta^{ab}+\frac{\rho^4}{3M^2}\Big[M^4(\delta^{ab}-e^ae^b)+2M\big(M^{ab}+\delta^{ab}M_{cd}e^ce^d-2e^{(a}M^{b)c}e_c\big)\\
&& \hspace{2cm}+2\big(S^aS^b+\delta^{ab}(S_ce^c)^2-2e^{(a}S^{b)}S_ce^c\big)\Big]+O(\rho^5).
\end{eqnarray*}
Notice that the metric $\bar{g}_{\mu\nu}$ is singular at
$\rho=0$. Thus, the points for which $\rho=0$ are at an infinite
distance with respect to this metric ---hence one does not obtain a
finite representation of spatial infinity. However, this singular behaviour is
 counteracted by the use of components with respect to the frame and
coframe basis introduced in the previous subsection. 

\medskip
It is also noticed that the conformal factor $\bar{\Omega}$ has the
following expansion:
\[
 \bar{\Omega}=\frac{1}{\rho}V^{1/2}\Omega=\rho+M\rho^2+\frac{1}{6M^2}\rho^3\big(5M^4+2MM_{ab}e^ae^b+2(S_ae^a)^2\big)+O(\rho^4).
\]
In the sequel, we will also require the components of the metric
$\bar{g}$ with respect to the frame $v_{\bm\alpha}$:
\begin{eqnarray*}
 &&\bar{g}_{\bm\alpha\bm\beta} =  \langle\Phi^*(\bar{g});v_{\bm\alpha},v_{\bm\beta}\rangle \\
 && \phantom{\bar{g}_{\bm\alpha\bm\beta}}= \langle(\bar{g}_{\mu\nu}\circ\Phi)\mbox{d}x^\mu \mbox{d}x^\nu;v_{\bm\alpha},v_{\bm\beta}\rangle \\
 & &\phantom{\bar{g}_{\bm\alpha\bm\beta}} =  (\bar{g}_{tt}\circ\Phi)\langle \mbox{d}t,v_{\bm\alpha}\rangle\langle \mbox{d}t,v_{\bm\beta}\rangle+2(\bar{g}_{ta}\circ\Phi)\langle \mbox{d}t,v_{\bm\alpha}\rangle\langle \mbox{d}x^a,v_{\bm\beta}\rangle \\
 && \hspace{2cm}+(\bar{g}_{ab}\circ\Phi)\langle \mbox{d}x^a,v_{\bm\alpha}\rangle\langle \mbox{d}x^b,v_{\bm\beta}\rangle \\
  && \phantom{\bar{g}_{\bm\alpha\bm\beta}}= (\bar{g}_{tt}\circ\Phi)v^t{}_{\bm\alpha} v^t{}_{\bm\beta}+2(\bar{g}_{ta}\circ\Phi)v^t{}_{\bm\alpha} v^a{}_{\bm\beta}+(\bar{g}_{ab}\circ\Phi)v^a{}_{\bm\alpha} v^b{}_{\bm\beta}.
\end{eqnarray*}
These components are explicitly given by 
\begin{eqnarray*}
&& \bar{g}_{\bm{00}}=0,\\
&& \bar{g}_{\bm{01}}=\frac{(V\Omega)[\rho\,e^a]}{(1-\bar{\tau})^2(V\Omega)[\bar{\rho}e^a]},\\
&&
\bar{g}_{\bm{0A}}=\frac{(V\Omega)[\rho\,e^a]}{(1-\bar{\tau})^2\bar{\rho}}\big(l_{\bm
  A}+\rho\,\beta_{\bm
  A}\big),\\
&& \bar{g}_{\bm{11}}=\frac{(V\Omega)[\rho\,e^a]}{(1-\bar{\tau})^2(V\Omega)[\bar{\rho}e^a]}\Bigg(\frac{(V\Omega)[\rho\,e^a]}{(V\Omega)[\bar{\rho}e^a]}-2(1-\bar{\tau})\Bigg),\\
&&
\bar{g}_{\bm{1A}}=\frac{(V\Omega)[\rho\,e^a]}{(1-\bar{\tau})^2\bar{\rho}}\Bigg(\frac{(V\Omega)[\rho\,e^a]}{(V\Omega)[\bar{\rho}e^a]}-(1-\bar{\tau})\Bigg)\big(l_A+\rho\,\beta_{\bm
  A}\big),\\
&&
\bar{g}_{\bm{AB}}=\frac{(V\Omega)^2[\rho\,e^a]}{(1-\bar{\tau})^2\bar{\rho}^2}\big(l_{\bm
  A}l_{\bm B}+2\rho\,\beta_{\bm A}l_{\bm B}\big)+k_{\bm
  A\bm B},
\end{eqnarray*}
where $\beta_{\bm A}=\beta_a\,e^a{}_{,\psi^A}$.

\medskip
The expansions discussed in the previous paragraphs imply that 
\[
\bar{g}_{\bm\alpha\bm\beta}=g^*_{\bm\alpha\bm\beta}+O(\bar{\rho}^2),
\]
with
\[
 (g^*_{\bm\alpha\bm\beta})=\left(\begin{array}{ccc}
                 0 & 1-2M\bar{\rho}\bar{\tau} & 0\\
                 1-2M\bar{\rho}\bar{\tau} & -(1-\bar{\tau})(1+\bar{\tau}-4M\bar{\rho}\bar{\tau}^2) & 0\\
                 0 & 0 & k_{\bm A \bm B}(0)
                \end{array}\right).
\]
In order to obtain these expressions we have used that
$k_{\bm A \bm B}(0)=\delta_{ab}\,e^a{}_{,\psi^A}\,e^b{}_{,\psi^B}$,
$\delta_{ab}\,e^a\,e^b=1$ and $\delta_{ab}\,e^a\,e^b{}_{,\psi^A}=0$. The
inverse metric is given by 
\[
\bar{g}^{\bm \alpha\bm\beta}={g^*}^{\bm\alpha\bm\beta}+O(\bar{\rho}^2),
\]
 with
\[
 ({g^*}^{\bm \alpha \bm\beta})=\left(\begin{array}{ccc}
                \displaystyle \frac{(1-\bar{\tau})(1+\bar{\tau}-4M\bar{\rho}\bar{\tau}^2)}{(1-2M\bar{\rho}\bar{\tau})^2} & \displaystyle\frac{1}{1-2M\bar{\rho}\bar{\tau}} & 0\\
                 \displaystyle\frac{1}{1-2M\bar{\rho}\bar{\tau}} & 0 & 0\\
                 0 & 0 & k^{\bm A \bm B}(0)
                \end{array}\right).
\]

\medskip
\noindent
\textbf{Remark 1.} One sees that although the conformal metric
$\bar{g}_{\mu\nu}$ is singular at $\rho=0$, its components measured
with respect to the frame of Subsection \ref{Subsection:Frame} are
regular and indicate the existence of an extended set with the
topology of a cylinder at spatial infinity ---its sections of constant
$\bar{\tau}$ correspond to 2-spheres.

\medskip
\noindent
\textbf{Remark 2.}
The stationary metric $\tilde{g}$ possesses the Killing vector
$\xi=\partial_t$. As the conformal factor $\Omega$ and $\rho$ do not
depend on $t$, $\xi$ is also a Killing vector of $\bar{g}$. In the new
coordinates it takes the form
\begin{eqnarray*}
 &&
 \xi=\frac{(V\Omega)[\bar{\rho}\,e^a]}{\bar{\rho}}\big((1-\bar{\tau})\partial_{\bar{\tau}}+\bar{\rho}\partial_{\bar{\rho}}\big)
 \\
&&\phantom{\xi}=\frac{(V\Omega)[\bar{\rho}\,e^a]}{\bar{\rho}}\big((1-\bar{\tau})v_{\bm
  0}+v_{\bm 1}\big).
\end{eqnarray*}

\medskip
\noindent
\textbf{Remark 3.}
The metric $\bar{g}$ is a conformal representation of the metric
$\breve{g}$ which allows us to construct an extension of $\cal{M}$ in a
neighbourhood of spatial infinity. For this, we replace the
hypersurface $\c{S}$ by the manifold with boundary $\bar{\cal S}$, by
adding to $\tc{S}$ the 2-dimensional surface $\partial\bar{\cal
  S}$. The points of this 2-dimensional surface are thought of as
ideal end points of radial curves on $\tc{S}$ as
$\bar{\rho}\rightarrow 0$. The coordinates $\bar{\rho}$, $\psi^A$
extend by definition to smooth coordinates on $\bar{\cal S}$ and on
$\partial\bar{\cal S}$ we have $\bar{\rho}=0$.  The coordinates
$\psi^A$, although not specified, are supposed to cover the
$\Sphere^2$. The construction described in this paragraph provides an alternative implementation of the
blow up of spatial infinity discussed in Subsection
\ref{Subsection:IntroCylinder} of the Introduction.

\medskip
Working by analogy with the discussion of Section \ref{regularProblem}, one defines the following regions in terms of the range of the coordinates $\bar{\tau}$, $\bar{\rho}$ and $\psi^A$:
\begin{eqnarray*}
&& {\tc{M}}'  \equiv  \left\{0\leq\bar{\tau}<1,0<\bar{\rho}\right\},\\
&& \bar{\c{M}}' \equiv  \left\{0\leq\bar{\tau}\leq1,0\leq\bar{\rho}\right\},\\
&& {\scri^{+}}' \equiv \left\{\bar{\tau}=1,\bar{\rho}>0\right\},\\
&& {\cal I}' \equiv \left\{0\leq\bar{\tau}<1,\bar{\rho}=0\right\},\\
&& {{\cal I}^+}' \equiv \left\{\bar{\tau}=1,\bar{\rho}=0\right\},\\
&& {{\cal I}^0}' \equiv \partial\bar{\cal S} = \left\{\bar{\tau}=0,\bar{\rho}=0\right\},\\
&& {\bar{\cal I}}' \equiv {\cal I}' \cup {{\cal I}^+}'.
\end{eqnarray*}

The names for these sets have been chosen in accordance with the
related sets defined in Section \ref{regularProblem}, although they
differ in some aspects. One can readily verify that in terms of the
new coordinates, frame and coframe fields, the metric $\bar{g}$
extends smoothly through the sets ${\scri^{+}}'$ and ${\bar{\cal
I}}'$. Accordingly, $\bar{\c{M}}'$ provides a suitable extension of
$\cal{M}$. In order to use this extension to show that the
construction of the cylinder at infinity for stationary spacetimes is
as smooth as expected we need also information regarding the Schouten
tensor and the conformal Weyl tensor, which we derive in the following
subsections. Finally, notice that in terms of the coordinates
$(\bar{\tau},\bar{\rho},\psi^A)$, null infinity appears to be parallel
to the surfaces of constant $\bar{\tau}$, and in particular the
initial hypersurface $\bar{\mathcal{S}}$.

\subsection{Expansions of the Schouten tensor}
\label{Schouten}

In this section we discuss expansions of the Schouten tensor of the
metric $\bar{g}$. This is related to the Schouten tensor of the metric $\tilde{g}$ by
\[
 \bar{L}_{\mu\nu}=\tilde{L}_{\mu\nu}-\frac{1}{\bar{\Omega}}\,\bar{\nabla}_\mu\bar{\nabla}_\nu\bar{\Omega}+\frac{1}{2\bar{\Omega}^2}\,\bar{\nabla}^\lambda\bar{\Omega}\,\bar{\nabla}_\lambda\bar{\Omega}\,\bar{g}_{\mu\nu}.
\]
In what follows, we will consider the components of $\bar{L}$ in the frame $v_{\bm\alpha}$:
\begin{eqnarray*}
&&  \bar{L}_{\bm \alpha\bm\beta} =  \langle\Phi^*(\bar{L});v_{\bm\alpha},v_{\bm\beta}\rangle \\
&&  \phantom{\bar{L}_{\alpha\beta}}= (\bar{L}_{tt}\circ\Phi)v^t{}_{\bm\alpha} v^t{}_{\bm\beta}+2(\bar{L}_{ta}\circ\Phi)v^t{}_{\bm\alpha} v^a{}_{\bm\beta}+(\bar{L}_{ab}\circ\Phi)v^a{}_{\bm\alpha} v^b{}_{\bm\beta}.
\end{eqnarray*}
As $\tilde{g}$ is a solution to the vacuum Einstein field equation, it
follows that $\tilde{L}_{\mu\nu}=0$. Furthermore, in view that
$\bar{\Omega}$ does not depend on $t$ one obtains
\[
 \bar{L}_{\mu\nu}=-\frac{1}{\bar{\Omega}}(\partial_\mu\partial_\nu\bar{\Omega}-\bar{\Gamma}_\mu{}^a{}_\nu\,\partial_a\bar{\Omega})+\frac{1}{2\bar{\Omega}^2}\,\bar{g}^{ab}\,\partial_a\bar{\Omega}\,\partial_b\bar{\Omega}\,\bar{g}_{\mu\nu}.
\]
Alternatively, one can write
\begin{eqnarray*}
&& \bar{L}_{tt} =  \frac{1}{\bar{\Omega}}\bar{\Gamma}_t\,^a\,_t\,\partial_a\bar{\Omega}+\frac{1}{2\bar{\Omega}^2}\,\bar{g}^{ab}\,\partial_a\bar{\Omega}\,\partial_b\bar{\Omega}\,\bar{g}_{tt},\\
&& \bar{L}_{ta} =  \frac{1}{\bar{\Omega}}\bar{\Gamma}_t\,^b\,_a\,\partial_b\bar{\Omega}+\frac{1}{2\bar{\Omega}^2}\,\bar{g}^{bc}\,\partial_b\bar{\Omega}\,\partial_c\bar{\Omega}\,\bar{g}_{ta},\\
&& \bar{L}_{ab}  =  -\frac{1}{\bar{\Omega}}\big(\partial_a\partial_b\bar{\Omega}-\bar{\Gamma}_a\,^c\,_b\,\partial_c\bar{\Omega}\big)+\frac{1}{2\bar{\Omega}^2}\,\bar{g}^{cd}\,\partial_c\bar{\Omega}\,\partial_d\bar{\Omega}\,\bar{g}_{ab}.
\end{eqnarray*}
Using the expansions for the conformal factor, the first and second
derivatives of the conformal factor and the Christoffel symbols of the
conformal metric given in Appendix \ref{expansions} ones get that
\begin{eqnarray*}
&&  \bar{L}_{tt} =  \frac{1}{2}\rho^2+4M\rho^3+O(\rho^4),\\
&& \bar{L}_{ta} =  -2\rho^2e_{abc}e^bS^c+O(\rho^3),\\
&& \bar{L}_{ab}  =  \frac{1}{2\rho^2}(\delta_{ab}-2e_ae_b)+\frac{M}{\rho}(\delta_{ab}-3e_ae_b)+O(\rho^0).
\end{eqnarray*}
One then has all the ingredients to calculate the components
$\bar{L}_{\bm\alpha\bm\beta}$. One obtains
\begin{eqnarray*}
&&\bar{L}_{\bm0\bm0} =  O(\bar{\rho}^2),\\
&&\bar{L}_{\bm0\bm1} =  \frac{1}{2}+M(2-3\bar{\tau})\bar{\rho}+O(\bar{\rho}^2),\\
&&\bar{L}_{\bm0 \bm A}  =  O(\bar{\rho}^2),\\
&&\bar{L}_{\bm1 \bm1}  =  -\frac{1}{2}(1-\bar{\tau}^2)-2M(1+2\bar{\tau}^2)(1-\bar{\tau})\bar{\rho}+O(\bar{\rho}^2),\\
&&\bar{L}_{\bm1\bm A}  =  O(\bar{\rho}^2),\\
&&\bar{L}_{\bm A\bm B}  =  \frac{1}{2}k_{\bm A \bm
  B}+M(1-\bar{\tau})\bar{\rho}\,k_{\bm A \bm B}+O(\bar{\rho}^2),
\end{eqnarray*}
where it has been used used that
\[
 \delta_{ab}\,e^a\!_{,\psi^A}\,e^b\!_{,\psi^B}=k_{AB}(0)+O(\bar{\rho}^2),
 \quad \delta_{ab}\,e^a\,e^b\!_{,\psi_A}=0.
\]
Summarising, one has proven the following result:

\begin{lemma}
The components, $\bar{L}_{\bm \alpha \bm \beta}$, of the Schouten tensor
$\bar{L}_{\mu\nu}$ in the frame $v_{\bm\alpha}$ are regular (i.e. non-divergent) at $\mathcal{I}'$.
\end{lemma}

\subsection{The Weyl tensor}

In this section we verify that conformal Weyl tensor has a smooth
limit at $\bar{\rho}\rightarrow 0$. For this, we start by recalling
the decomposition of the Weyl tensor in terms of its electric and
magnetic parts with respect to the normal to a hypersurface.

\subsubsection{The electric-magnetic decomposition}
 Let $\tc{M}$ be the spacetime with metric $\t{g}_{\mu\nu}$ and $\tc{S}$ a
 space-like hypersurface with unit normal $\t{n}^\mu$. The induced metric
 on $\tc{S}$ by $\t{g}_{\mu\nu}$ is given by
\[
\t{h}_{\mu\nu}=\t{g}_{\mu\nu}-\t{n}_\mu\t{n}_\nu.
\]
For convenience one defines 
\begin{equation}
\t{p}_{\mu\nu}\equiv \t{h}_{\mu\nu}-\t{n}_\mu\t{n}_\nu, \quad \t{\epsilon}_{\mu\nu\lambda}\equiv\t{n}^\rho\t{\epsilon}_{\rho\mu\nu\lambda}. 
\label{Definition:p}
\end{equation}
The $\t{n}$-electric and $\t{n}$-magnetic parts of the conformal Weyl
tensor are given, respectively, by
\begin{equation}
\label{partsC}
 \t{c}_{\nu\rho}\equiv\t{C}_{\mu\nu\lambda\rho}\t{n}^\mu\t{n}^\lambda,\qquad
 \t{c}^*_{\nu\rho}\equiv \t{C}^*_{\mu\nu\lambda\rho}\t{n}^\mu\t{n}^\lambda,
\end{equation}
where $\t{C}^*_{\mu\nu\lambda\rho}$ denotes the dual of the conformal
Weyl tensor:
$\t{C}^*_{\mu\nu\lambda\rho}=\frac{1}{2}\t{C}_{\mu\nu\sigma\chi}\t{\epsilon}^{\sigma\chi}\,_{\lambda\rho}$.The
$\t{n}$-electric and $\t{n}$-magnetic parts of the Weyl tensor are
symmetric, trace-free and spatial:
\[
 \t{n}^\mu\t{c}_{\mu\nu}=0, \qquad \t{n}^\mu\t{c}^*_{\mu\nu}=0.
\]
 Given two tensors of rank two, $f_{\mu\nu}$ and $k_{\mu\nu}$, their \emph{Kulkarni-Nomizu} product is defined as
\[
 (f \oslash k)_{\mu\nu\lambda\rho}=2(f_{\mu[\lambda}k_{\rho]\nu}-f_{\nu[\lambda}k_{\rho]\mu}).
\]
In terms of the Kulkarni-Nomizu product, the conformal Weyl tensor is given by
\begin{eqnarray*}
&& \t{C}_{\mu\nu\lambda\rho} =  2\big(\t{p}_{\nu[\lambda}\t{c}_{\rho]\mu}-\t{p}_{\mu[\lambda}\t{c}_{\rho]\nu}-\t{n}_{[\lambda}\t{c}^*_{\rho]\delta}\t{\epsilon}^\delta\,_{\mu\nu}-\t{n}_{[\mu}\t{c}^*_{\nu]\delta}\t{\epsilon}^\delta\,_{\lambda\rho}\big)\\
&&  \phantom{ \t{C}_{\mu\nu\lambda\rho}}=  (\t{p}\oslash\t{c}^*)^*_{\mu\nu\lambda\rho}-(\t{p}\oslash\t{c})_{\mu\nu\lambda\rho}.
\end{eqnarray*}

\subsubsection{Conformal rescalings}
If $\t{g}_{\mu\nu}$ is a solution of the Einstein vacuum field equations, then
the first and second fundamental forms of $\tc{S}$ induced by $\t{g}_{\mu\nu}$
satisfy the Gauss and the Codazzi equations. These equations allow to
write the pull-back of $\t{c}_{\mu\nu}$ and $\t{c}^*_{\mu\nu}$ to
$\tc{S}$ in terms of the initial data quantities. More precisely,
\begin{equation}\label{Geq}
 \t{c}_{ab}=-r_{ab}[\t{h}]+\t{\chi}_c\,^c\t{\chi}_{ab}-\t{\chi}_{ca}\t{\chi}_b\,^c,
 \qquad \t{c}^*_{ab}=-\t{D}_c\t{\chi}_{d(a}\t{\epsilon}_{b)}\,^{cd}.
\end{equation}
If $\bar{g}_{\mu\nu}=\bar{\Omega}^2\t{g}_{\mu\nu}$, then it is well
known that 
$\bar{C}^\mu\,_{\nu\lambda\rho}=\tilde{C}^\mu\,_{\nu\lambda\rho}$. The
rescaled conformal Weyl tensor is given by
$\bar{W}^\mu\,_{\nu\lambda\rho}=\bar{\Omega}^{-1}\bar{C}^\mu\,_{\nu\lambda\rho}$,
and therefore
\[
 \bar{W}_{\mu\nu\lambda\rho}=\bar{\Omega}\t{C}_{\mu\nu\lambda\rho}.
\]
The $\bar{n}$-electric and $\bar{n}$-magnetic parts of $\bar{W}$ are
defined in accordance with \eqref{partsC} as
\[
 \bar{w}_{\nu\rho}=\bar{W}_{\mu\nu\lambda\rho}\bar{n}^\mu\bar{n}^\lambda,\qquad\bar{w}^*_{\nu\rho}=\bar{W}^*_{\mu\nu\lambda\rho}\bar{n}^\mu\bar{n}^\lambda.
\]
Now, recalling that the $\bar{g}$-unit normal to $\tc{S}$ is given by  $\bar{n}^{\mu}=\bar{\Omega}^{-1}\t{n}^\mu$ one obtains
\begin{equation}\label{ctow}
 \bar{w}_{\mu\nu}=\bar{\Omega}^{-1}\t{c}_{\mu\nu},\hspace{2cm}\bar{w}^*_{\mu\nu}=\bar{\Omega}^{-1}\t{c}^*_{\mu\nu}.
\end{equation}
From the definition \eqref{Definition:p} one readily obtains
\begin{eqnarray*}
 && \bar{p}_{\mu\nu}=\bar{\Omega}^2\t{p}_{\mu\nu}, \\
 &&\bar{W}_{\mu\nu\lambda\rho} =  (\bar{p}\oslash\bar{w}^*)^*_{\mu\nu\lambda\rho}-(\bar{p}\oslash\bar{w})_{\mu\nu\lambda\rho}\\
&& \phantom{\bar{W}_{\mu\nu\lambda\rho}} =  2\big(\bar{p}_{\nu[\lambda}\bar{w}_{\rho]\mu}-\bar{p}_{\mu[\lambda}\bar{w}_{\rho]\nu}-\bar{n}_{[\lambda}\bar{w}^*_{\rho]\delta}\bar{\epsilon}^\delta\,_{\mu\nu}-\bar{n}_{[\mu}\bar{w}^*_{\nu]\delta}\bar{\epsilon}^\delta\,_{\lambda\rho}\big).
\end{eqnarray*}

\subsubsection{Regularity at $\mathcal{I}'$}
In what follows, it will be shown that the components of the Weyl
tensor $\bar{W}_{\mu\nu\lambda\rho}$ with respect to the frame
$v_{\bm\alpha}$ given by
\begin{eqnarray*}
 && \bar{W}_{\bm\alpha\bm\beta\bm\gamma\bm\delta} = \langle\Phi^*(\bar{W});v_{\bm\alpha},v_{\bm\beta},v_{\bm\gamma},v_{\bm\delta}\rangle \\
&& \phantom{\bar{W}_{\alpha\beta\gamma\delta}}=(\bar{W}_{\mu\nu\lambda\rho}\circ\Phi)v^\mu{}_{\bm\alpha} v^\nu{}_{\bm\beta} v^\lambda{}_{\bm\gamma} v^\rho{}_{\bm\delta}
\end{eqnarray*}
do not diverge as $\bar{\rho}\rightarrow 0$. In order to do this, one
needs to discuss the expansions of $\bar{n}_\mu$, $\bar{\epsilon}^\mu\,_{\nu\lambda}$, $\bar{p}_{\mu\nu}$, $\bar{w}_{\mu\nu}$ and $\bar{w}^*_{\mu\nu}$.

\medskip
We notice that the hypersurface to be considered is given in our coordinates by
\[
 \tc{S}=\{\phi(x^\mu)=t-constant=0\}.
\]
It follows that the normal vector and covector are given by
\begin{eqnarray*}
\bar{n}_t=\frac{V\Omega}{\rho(1-V^2\Omega^2\beta_c\beta^c)^{1/2}}, & \hspace{1cm} & \bar{n}_a=0,\\
\bar{n}^t=\frac{\rho(1-V^2\Omega^2\beta_c\beta^c)^{1/2}}{V\Omega}, & \hspace{1cm} & \bar{n}^a=\frac{\rho V\Omega\beta^a}{(1-V^2\Omega^2\beta_c\beta^c)^{1/2}},
\end{eqnarray*}
whence
\begin{eqnarray*}
&&\bar{n}_t =  \rho+2M\rho^2+\frac{1}{3M^2}\rho^3\big(7M^4+MM_{ab}e^ae^b+(S_ae^a)^2\big)+O(\rho^4),\\
&&\bar{n}_a =  0,\\
&&\bar{n}^t =  \rho^{-1}-2M+\frac{1}{3M^2}\rho\big(5M^4-MM_{ab}e^ae^b-(S_ae^a)^2\big)+O(\rho^4),\\
&&\bar{n}^a =  -2\rho^3e^{abc}e_bS_c+O(\rho^4).
\end{eqnarray*}
To compute $\bar{\epsilon}^\mu\,_{\nu\lambda}$ one notes the relations 
$\bar{\epsilon}^\mu\,_{\nu\lambda}=\bar{g}^{\mu\rho}\bar{\epsilon}_{\rho\nu\lambda}$
and that $\bar{\epsilon}_{\mu\nu\lambda}=\bar{n}^\sigma\bar{\epsilon}_{\sigma\mu\nu\lambda}$, where
\[
\bar{\epsilon}_{\sigma\mu\nu\lambda}=|\det(\bar{g}_{\mu\nu})|^\frac{1}{2}\eta_{\sigma\mu\nu\lambda}=\frac{1}{\rho^2}(1+2M\rho+O(\rho^2))\eta_{\sigma\mu\nu\lambda},
\]
with $\eta_{\sigma\mu\mu\lambda}$ the totally antisymmetric tensor of
rank 4. Next one
evaluates $\bar{p}_{\mu\nu}=\bar{g}_{\mu\nu}-2\bar{n}_\mu\bar{n}_\nu$
to obtain
\[
(\bar{p}_{\mu\nu}) = \left(
\begin{array}{c c}
 \displaystyle -\frac{V^2\Omega^2}{\rho^2}\frac{(1+V^2\Omega^2\beta_c\beta^c)}{(1-V^2\Omega^2\beta_c\beta^c)}
  &\displaystyle \frac{1}{\rho^2}V^2\Omega^2\beta_a \\
  \displaystyle\frac{1}{\rho^2}V^2\Omega^2\beta_b &
  \displaystyle \frac{1}{\rho^2}(-\gamma_{ab}+V^2\Omega^2\beta_a\beta_b) 
\end{array} \right),
\]
from which
\begin{eqnarray*}
&&\bar{p}_{tt} =  -\rho^2-4M\rho^3-\frac{2}{3M^2}\rho^4\big(13M^4+MM_{ab}e^ae^b+(S_ae^a)^2\big)+O(\rho^5),\\
&&\bar{p}_{ta}  =  -2\rho^2e_{abc}e^bS^c+O(\rho^3),\\
&&\bar{p}_{ab}  =  -\rho^{-2}\delta_{ab}-\frac{1}{3M^2}\Big(M^4(\delta_{ab}-e_ae_b)+2M\big(M_{ab}+\delta_{ab}M_{cd}e^ce^d-2e_{(a}M_{b)c}e^c\big)\\
&&\hspace{2cm}+2\big(S_aS_b+\delta_{ab}(S_ce^c)^2-2e_{(a}S_{b)}S_ce^c\big)\Big)+O(\rho).
\end{eqnarray*}

\medskip
In order to calculate expansions for the electric and magnetic parts
$\bar{w}_{\mu\nu}$ and $\bar{w}^*_{\mu\nu}$ one makes use of the
expressions \eqref{ctow} and \eqref{Geq}. From the  formulae for conformal
transformations one has 
\[
 \t{r}_{ab}=\bar{r}_{ab}+\bar{\Omega}^{-1}\bar{D}_a\bar{D}_b\bar{\Omega}+\bar{\Omega}^{-1}\bar{D}^c\bar{D}_c\bar{\Omega}\bar{h}_{ab}-2\bar{\Omega}^{-2}\bar{D}^c\bar{\Omega}\bar{D}_c\bar{\Omega}\bar{h}_{ab}
\]
and
\begin{equation}
 \bar{\chi}_{ab}=\bar{\Omega}(\t{\chi}_{ab}+\bar{\Sigma}\t{h}_{ab}),
\end{equation}
where $\bar{\Sigma}=\bar{n}^\mu\,\partial_\mu\bar{\Omega}$. Hence one obtains
\begin{eqnarray}
&& \bar{w}_{ab} =
-\bar{\Omega}^{-1}\Big(\bar{r}_{ab}+\bar{\Omega}^{-1}\bar{D}_a\bar{D}_b\bar{\Omega}+\bar{\Omega}^{-1}\bar{D}^c\bar{D}_c\bar{\Omega}\bar{h}_{ab}-\bar{\chi}_c\,^c\bar{\chi}_{ab}+\bar{\chi}_{ca}\bar{\chi}_b\,^c
\nonumber \\
 &&
 \hspace{3cm}+\bar{\Omega}^{-1}\bar{\Sigma}\bar{\chi}_{ab}+\bar{\Omega}^{-1}\bar{\Sigma}\bar{\chi}_c\,^c\bar{h}_{ab}-2\bar{\Omega}^{-2}\bar{D}^c\bar{\Omega}\bar{D}_c\bar{\Omega}\bar{h}_{ab}-2\bar{\Omega}^{-2}\bar{\Sigma}^2\bar{h}_{ab}\Big). \label{wbar:firstexpression}
\end{eqnarray}
The seemingly more singular terms in the last expression can be
eliminated using the Hamiltonian constraint
\begin{eqnarray*}
&&0 =  \t{r}-(\t{\chi}_c\,^c)^2+\t{\chi}_{cd}\t{\chi}^{cd}\\
&& \phantom{0}=  \bar{\Omega}^2\bar{r}+4\bar{\Omega}\bar{D}^c\bar{D}_c\bar{\Omega}-6\bar{D}^c\bar{\Omega}\bar{D}_c\bar{\Omega}
 -\bar{\Omega}^2(\bar{\chi}_c\,^c)^2+\bar{\Omega}^2\bar{\chi}_{cd}\bar{\chi}^{cd}+4\bar{\Omega}\bar{\Sigma}\bar{\chi}_c\,^c-6\bar{\Sigma}^2,
\end{eqnarray*}
so that equation \eqref{wbar:firstexpression} transforms into
\begin{eqnarray}
&&\bar{w}_{ab} =
-\bar{\Omega}^{-1}\Big(\bar{r}_{ab}-\tfrac{1}{3}\bar{r}\bar{h}_{ab}+\bar{\Omega}^{-1}\big(\bar{D}_a\bar{D}_b\bar{\Omega}-\tfrac{1}{3}\bar{D}^c\bar{D}_c\bar{\Omega}\bar{h}_{ab}\big)
\nonumber \\
&&
\hspace{3cm}-(\bar{\chi}_c\,^c-\bar{\Omega}^{-1}\bar{\Sigma})\big(\bar{\chi}_{ab}-\tfrac{1}{3}\bar{\chi}_c\,^c\bar{h}_{ab}\big)+\bar{\chi}_{ca}\bar{\chi}_b\,^c-\tfrac{1}{3}\bar{\chi}_{cd}\bar{\chi}^{cd}\bar{h}_{ab}\Big). \label{elecW}
\end{eqnarray}
To calculate $\bar{w}^*_{ab}$ one needs to use the conformal transformation for the derivative of a $(0,2)$-tensor:
\[
 \t{D}_a\t{\chi}_{bc}=\bar{D}_a\t{\chi}_{bc}+\bar{\Omega}^{-1}\big(2\partial_a\bar{\Omega}\t{\chi}_{bc}+\partial_b\bar{\Omega}\t{\chi}_{ac}+\partial_c\bar{\Omega}\t{\chi}_{ba}-\bar{h}_{ab}\bar{h}^{de}\partial_e\bar{\Omega}\t{\chi}_{dc}-\bar{h}_{ac}\bar{h}^{de}\partial_d\bar{\Omega}\t{\chi}_{db}\big).
\]
The latter, together with
$\bar{\epsilon}_{abc}=\bar{\Omega}^3\t{\epsilon}_{abc}$ finally yield
\begin{equation}\label{barw}
 \bar{w}^*_{ab}=-\bar{\Omega}^{-1}\bar{D}_c\bar{\chi}_{d(a}\bar{\epsilon}_{b)}\,^{cd}.
\end{equation}
In order to compute $\bar{w}_{\mu\nu}$ and $\bar{w}^*_{\mu\nu}$ from
the expressions \eqref{elecW} and \eqref{barw} one uses that
$\t{n}^\mu\t{c}_{\mu\nu}=0$ and $\t{n}^\mu\t{c}^*_{\mu\nu}=0$ so that
$\bar{n}^\mu\bar{w}_{\mu\nu}=0$ and
$\bar{n}^\mu\bar{w}^*_{\mu\nu}=0$. With this and the symmetry of the
tensors one obtains
\begin{eqnarray*}
\bar{w}_{tt}=\frac{\bar{n}^a\bar{n}^b\bar{w}_{ab}}{(\bar{n}^t)^2},& \hspace{1cm}&\bar{w}_{ta}=-\frac{\bar{n}^b\bar{w}_{ba}}{\bar{n}^t},\\
\bar{w}^*_{tt}=\frac{\bar{n}^a\bar{n}^b\bar{w}^*_{ab}}{(\bar{n}^t)^2},&\hspace{1cm}&\bar{w}^*_{ta}=-\frac{\bar{n}^b\bar{w}^*_{ba}}{\bar{n}^t}.
\end{eqnarray*}

\medskip
In order to evaluate the formula \eqref{elecW} one makes use of 
\begin{eqnarray*}
&&\bar{r}_{ab} =  \partial_c\bar{\Gamma}_a{}^c{}_b-\partial_a\bar{\Gamma}_c{}^c{}_b+\bar{\Gamma}_a{}^c{}_b\,\bar{\Gamma}_c{}^d{}_d-\bar{\Gamma}_c{}^d{}_b\,\bar{\Gamma}_d{}^c{}_a\\
&& \phantom{\bar{r}_{ab}}=  \frac{1}{\rho^2}(\delta_{ab}-e_ae_b)-\frac{2}{3M^2}\Big(M^4(\delta_{ab}+e_ae_b)+M(M_{ab}+2\delta_{ab}M_c\,^c-3\delta_{ab}M_{cd}e^ce^d\\
&& \hspace{2cm}+4e_{(a}M_{b)c}e^c+S_aS_b+2\delta_{ab}S_cS^c-3\delta_{ab}(S_ce^c)^2+4e_{(a}S_{b)}S_ce^c\Big)+O(\rho),
\end{eqnarray*}
and
\begin{eqnarray*}
&& \bar{D}_a\bar{D}_b\bar{\Omega} =  \partial_a\partial_b\bar{\Omega}-\bar{\Gamma}_a{}^c{}_b\,\partial_c\bar{\Omega}\\
&& \phantom{\bar{D}_a\bar{D}_b\bar{\Omega}}=
\frac{1}{\rho}e_ae_b+4Me_ae_b+\frac{1}{3M^2}\rho\Big( 4e_{(a}S_{b)}S_ce^c+e_ae_b(S_ce^c)^2+ M^4\left(\delta_{ab}+\tfrac{43}{2}e_ae_b\right)\\
&&\hspace{2cm}+M\big(4M_{ab}+e_ae_bM_{cd}e^ce^d+4e_{(a}M_{b)c}e^c\big)+4S_aS_b\Big)+O(\rho^2).
\end{eqnarray*}
To complete the analysis, one also needs expansions for
$\bar{\chi}_{ab}$. To this end we make use of the tensor $\breve{\chi}_{ab}$ and
the results in \cite{Dai01b}. More precisely, one has that 
\[
 \breve{\chi}_{ab}=\breve{\Omega}^2\breve{\psi}_{ab},\qquad \breve{\psi}_{ab}=\frac{3}{\rho^3}e_{(a}e_{b)cd}S^ce^d+O(\rho^{-2}),
\]
so that
\[
 \breve{\chi}_{ab}=3\rho e_{(a}e_{b)cd}S^ce^d+O(\rho^2).
\]
To get from $\breve{\chi}_{ab}$ to $\bar{\chi}_{ab}$ one uses the
corresponding conformal transformation formulae to find that
\[
\bar{\chi}_{ab}=3e_{(a}e_{b)cd}S^ce^d+O(\rho).
\]
Finally, it is noticed that
\[
\bar{\Sigma}=\bar{n}^\mu\partial_\mu\bar{\Omega}=O(\rho^4), \qquad \bar{\epsilon}_a\,^{bc}=\rho e_a\,^{bc}+O(\rho^2)
\]
It follows then from formula \eqref{elecW} that
\begin{eqnarray*}
&&\bar{w}_{ab} =  \rho^{-2}M(\delta_{ab}-3e_ae_b)+O(\rho^0),\\
&&\bar{w}_{ta} =  2M\rho^2 e_{abc}e^bS^c+O(\rho^3)=-M\rho^2\beta_a+O(\rho^3),\\
&&\bar{w}_{tt}  =  4M\rho^6(S_aS^a-(S_ae^a)^2)+O(\rho^7).
\end{eqnarray*}
Similarly, one finds from \eqref{barw} that
\begin{eqnarray*}
&&\bar{w}^*_{ab}  =  \tfrac{3}{2}\rho^{-1}\left((\delta_{ab}+e_ae_b)S_ce^c-4e_{(a}S_{b)}\right)+O(\rho^0),\\
&&\bar{w}^*_{ta}  =  3\rho^3e_{abc}e^bS^cS_de^d+O(\rho^4),\\
&&\bar{w}^*_{tt}  =  -6\rho^7(S_aS^a-(S_ae^a)^2)S_be^b+O(\rho^8).
\end{eqnarray*}

\medskip
In view of the previous discussion, one has all the ingredients to
compute the leading terms of the expansions of the components
$\bar{W}_{\bm{\alpha\beta\gamma\delta}}$. Due to the length of the
calculation this has been done in a tensor manipulation
program. One obtains the following:

\begin{lemma}
The components, $\bar{W}_{\bm \alpha\bm\beta\bm\gamma\bm\delta}$, of the Weyl tensor
$\bar{W}_{\mu\nu\lambda\rho}$ in the frame $v_{\bm\alpha}$ are regular (i.e. non-divergent) at $\mathcal{I}'$.
\end{lemma}

\section{Stationary vacuum solutions near the cylinder at space-like
  infinity}
\label{Section:CylinderStationary}

Once the regularity of the various conformal field at $\mathcal{I}'$
has been shown, the last step in our analysis is very similar to the
discussion in \cite{Fri04}. The proof consists of several parts: first,
one starts by giving explicitly a solution to the conformal geodesic
equations on $\bar{\cal I}'$; in a second step a stability argument is
used to show that the construction of the cylinder at spacelike
infinity is regular in a neighbourhood of $\bar{\cal I}$; finally, one
needs to show that the whole construction does not depend on the
choice of conformal factor on the initial hypersurface.

\medskip
\noindent
\textbf{Remark.} It is important to stress the differences in the
regularity of static and stationary fields at spatial infinity. In the
static case all relevant fields are analytic. By contrast, as shown in
\cite{Dai01b}, in the strictly stationary case the relevant fields are
never analytic as functions of asymptotically Cartesian
coordinates. However, as already seen, the stationary fields have an
analytic expansion in terms of radial and angular coordinates. This is
the type of coordinates used in both the general construction of the
cylinder at spacelike infinity and in the extension discussed in
Section \ref{confExtSVS}. As a
consequence, all the relevant fields are analytic in these
coordinates.

\medskip
\emph{In what follows, the word analyticity will be used to
describe analytic behaviour with respect to the radial and angular coordinates.}

\subsection{Setting the conformal Gauss system} 

We consider now  the regular finite initial value
problem at spacelike infinity for stationary data. For this, we
make use of the initial hypersurface
\[
\tc{S}=\{t=0\},
\]
and set the initial conditions on $\tc{S} $ that generate the
desired conformal Gauss gauge system.

\subsubsection{Initial data for the canonical conformal factor}
The initial data for the conformal factor, $\Theta_*$,
is prescribed ---cfr. equation \eqref{Thetastar}--- by means of the function  
\begin{equation}
\label{Choice:kappa}
\kappa\equiv 2\breve{\Omega}|\breve{D}_a\breve{\Omega}\breve{D}^a\breve{\Omega}|^{-1/2}
\end{equation}
so that
\[
\Theta_*=\kappa^{-1}\breve{\Omega}.
\]
It follows that
\begin{eqnarray*}
&&\kappa=\rho+O(\rho^2),\\
&&\Theta_*=\frac{1}{2}|\breve{D}_a\breve{\Omega}\breve{D}^a\breve{\Omega}|^{1/2}=\rho+O(\rho^2).
\end{eqnarray*}
Hence, the conformal metric evaluated on $\tc{S}$ and the induced
metric are given, respectively, by 
\[
g_{\mu\nu}=\Theta_*^2\t{g}_{\mu\nu}, \qquad h_{ab}=\Theta_*^2\t{h}_{ab}=\kappa^{-2}\breve{h}_{ab}.
\]

\subsubsection{Initial data for the tangent vector to the congruence
  of conformal geodesics}
Initial conditions for the tangent vector $\dot{x}=\mbox{d}x/\mbox{d}\tau$,
where $\tau$ is the parameter of the conformal
geodesic, are set to be 
\[
\dot{x}\perp\tc{S},\hspace{1cm}g(\dot{x},\dot{x})=1, \quad \mbox{ on }
\tc{S}.
\]
In order to implement the requirement of having $\dot{x}$ orthogonal to
$\tc{S}$ we consider the stationary Killing vector $\xi$. One has that
\[
\xi=\frac{V\Omega}{\bar{\rho}}(v_{\bm 0}+v_{\bm 1}) \quad \mbox{ on } \tc{S}, 
\]
where $v_{\bm0}$ and $v_{\bm 1}$ are the first two vectors of the frame
$v_{\bm\alpha}$ discussed in Subsection \ref{Subsection:Frame}. It can be
readily checked that
$\langle\xi,\mbox{d}\bar{\rho}\rangle=\langle\xi,\mbox{d}\psi^A\rangle=0$
on $\tc{S}$ so that $\xi\perp\tc{S}$. In order to obtain the right
normalisation one considers
\[
g(\xi,\xi)=\frac{\Theta^2}{\bar{\Omega}^2}\bar{g}(\xi,\xi)=\frac{\Theta^2}{\bar{\Omega}^2}\frac{V^2\Omega^2}{\bar{\rho}^2}.
\]
Thus,
\begin{equation}\label{inx}
\dot{x}=\frac{\bar{\Omega}}{\Theta}(v_{\bm 0}+v_{\bm
  1})=\frac{\kappa}{\bar{\rho}}(v_{\bm 0}+v_{\bm 1})\equiv X^{\bm
  \alpha}v_{\bm \alpha},
\end{equation}
where
\begin{equation}\label{inX}
X^{\bm\alpha}=\delta_0{}^{\bm \alpha}+\delta_1^{\bm \alpha}+O(\bar{\rho}).
\end{equation}

\subsubsection{Initial data for the 1-form $f$}
The initial data for the 1-form $f$ is chosen in agreement with
condition \eqref{initialf} so that
\[
\langle f,\partial_\tau\rangle=0,\quad \mbox{pull back of $f$ to $\tc{S}$}=\kappa^{-1}d\kappa.
\]
It follows then that $\langle f,\dot{x}\rangle=0$. This property,
together with the choice \eqref{Choice:kappa} for the function
$\kappa$  ---see equation \eqref{Thetasol}--- give
\begin{equation}\label{confF}
\Theta=\Theta_*\left(1-\tau^2\right).
\end{equation}

\subsection{Relating the various conformal gauges}

The analysis performed in Section \ref{confExtSVS} was done in terms
of the metric $\bar{g}_{\mu\nu}$. Now, we proceed to discuss how this metric
and its Levi-Civita connection, $\bar{\nabla}$, are related to the
metric $g_{\mu\nu}$ and its Levi-Civita  connection, $\nabla$. From the relation
$\bar{g}_{\mu\nu}=\bar{\Omega}^2\t{g}_{\mu\nu}$ it follows that
\[
g_{\mu\nu}=\Pi^2\bar{g}_{\mu\nu}=\Theta^2\t{g}_{\mu\nu},\hspace{1cm}\Pi\equiv\bar{\Omega}^{-1}\Theta.
\]
The relations between the physical Levi-Civita connection
$\t{\nabla}$, the unphysical Levi-Civita connections $\bar{\nabla}$,
$\nabla$, and the Weyl connection $\hat{\nabla}$ are given by
\begin{eqnarray*}
&&\hat{\nabla}=\t{\nabla}+S(\t{f}),\\
&&\hat{\nabla}=\nabla+S(f),\\
&&\hat{\nabla}=\bar{\nabla}+S(\bar{f}),\\
&&\nabla=\t{\nabla}+S(\Theta^{-1}\mbox{d}\Theta),\\
&&\bar{\nabla}=\t{\nabla}+S(\bar{\Omega}^{-1}\mbox{d}\bar{\Omega}), 
\end{eqnarray*}
where 
\[
\bar{f}=f+\Pi^{-1}\mbox{d}\Pi=f+\Theta^{-1}\mbox{d}\Theta-\bar{\Omega}^{-1}\mbox{d}\bar{\Omega}.
\]
From the particular form of the conformal factor $\Theta$ given by equation
\eqref{confF} one has that $\langle \mbox{d}\Theta,\dot{x}\rangle=0$
on $\tc{S}$. As $\bar{\Omega}$ is independent of $t$ and $\partial_t$
is orthogonal $\tc{S}$, it follows that $\langle
\mbox{d}\bar{\Omega},\dot{x}\rangle=0$. Thus,
$\langle\bar{f},\dot{x}\rangle=0$ on $\tc{S}$. Finally, from 
\begin{equation}
\label{eqPi}
\Pi=\frac{\bar{\rho}}{\kappa} \quad \mbox{ on } \tc{S},
\end{equation} 
and observing that the pull-back of $\bar{f}$ to $\tc{S}$ is given by
$\bar{\rho}^{-1}\mbox{d}\bar{\rho}$, it follows that
\begin{equation}
\label{inf}
\bar{f}=\bar{f}_{\bm \alpha}\alpha^{\bm \alpha},\quad \mbox{ with
}\quad \bar{f}_{\bm \alpha}=-\delta_{\bm \alpha}{}^0+\delta_{\bm \alpha}{}^1 \quad \mbox{ on }\tc{S}.
\end{equation}
A property of conformal Gaussian systems is that $\langle
f,\dot{x}\rangle=0$ in the whole of the spacetime. Hence, one has that
\begin{equation}
\label{Pidot}
\dot{\Pi}=\Pi\langle\bar{f},\dot{x}\rangle
\end{equation}
along the conformal geodesics. This last equation, together with
equation \eqref{eqPi}, allows to determine $\Pi$ if the
contraction $\langle\bar{f},\dot{x}\rangle$ is known.

\subsection{Solving the conformal geodesic equations}
In this section we provide a discussion of the conformal geodesic
equations with respect to the metric $\bar{g}$ and of its solutions.
A solution to these equations is given by a spacetime curve
$x(\tau)=(\bar{\tau}(\tau),\bar{\rho}(\tau),\psi^A(\tau))$ and a
$1$-form $\bar{f}(\tau)$ along the curve. If expressed in terms of the
frame fields, $v_{\bm \alpha}$, and coframe fields, $\alpha^{\bm \alpha}$, the functions involved
in the conformal geodesic equations are the components
$\bar{g}_{\bm\alpha\bm\beta}$, $\bar{g}^{\bm\alpha\bm\beta}$,
$\bar{\Gamma}_{\bm\alpha}{}^{\bm\gamma}{}_{\bm\beta}$ and
$\bar{L}_{\bm\alpha\bm\beta}$. These functions extend by analyticity
through ${\bar{\cal I}}'$ into a domain where $\bar{\rho}<0$. If one
assumes such an extension, one obtains the so-called {\it extended
conformal geodesic equations}. It is important to point out that the
initial data on $\bar{\cal S}$ are analytic. Therefore, one can
consider these equations in a neighbourhood of ${\bar{\cal
I}}'$. Moreover, it turns out that the restriction of the equations to
${{\cal I}^0}'$ can be solved explicitly. The solution one obtains is
universal in the sense that it is the same for all stationary
solutions with non-vanishing mass. More
precisely, one has the following lemma, whose proof is the same as that
of Lemma 7.2 in \cite{Fri04}:

\begin{lemma}
\label{Lemma:CGExplicitSolns}
The solution to the restriction of the extended conformal geodesic
equations to  ${\bar{\cal I}}'$ with the components
$\bar{L}_{\bm\alpha\bm\beta}$ as given in Subsection \ref{Schouten}
and  initial data $x=(0,0,{\psi^A}')$,
$X^{\bm\alpha}=\delta_0{}^{\bm\alpha}+\delta_1{}^{\bm\alpha}$ and
$\bar{f}_{\bm\alpha}=-\delta_{\bm\alpha}{}^0+\delta_{\bm\alpha}{}^1$ is given by
\begin{eqnarray*}
&& x(\tau)=(\bar{\tau}(\tau),0,\psi^A(\tau))=(\tau,0,{\psi^A}'),\\
&& \bar{f}_{\bm0}=-\frac{1}{1+\tau}, \quad \bar{f}_{\bm1}=1, \quad
\bar{f}_{\bm A}=0.
\end{eqnarray*}
This solution extends by analyticity to a domain
$0\leq\tau\leq1+2\epsilon$ for some $\epsilon>0$. The extension to
${\bar{\cal I}}'$ of the conformal factor $\Pi$  determined
by equations \eqref{Pidot} and \eqref{eqPi} takes on  ${\bar{\cal I}}'$ the value $\Pi=1$.
\end{lemma}

The previous lemma not only gives precise information about the
conformal geodesics ruling ${\bar{\cal I}}'$, but it also shows that
these geodesics extend analytically beyond $\tau=1$. These facts are,
in turn, used to show that there exists a solution of the conformal geodesic
equations near ${\bar{\cal I}}'$, and that this solution extends for
sufficiently large values of  the parameter $\tau$.

\medskip
Consider a smooth extension of $\tc{S}$ into a range where
$\bar{\rho}<0$ such that $(\bar{\rho},\psi^A)$ extend as smooth
coordinates. We denote this extension by $\bar{\cal S}_{ext}$. Now, if
the extension $\bar{\cal S}_{ext}\backslash\tc{S}$ is small enough,
then the initial conditions for the conformal geodesic
equations \eqref{inx}, \eqref{inX} and \eqref{inf} extend analytically
to $\bar{\cal S}_{ext}$ ---the precise range of $\bar{\rho}$ is not required as long as it is 
small enough. Therefore, the conformal geodesic equations determine near $\bar{\cal
  S}_{ext}$ an analytic congruence of solutions to the extended
conformal geodesic equations. From Lemma \ref{Lemma:CGExplicitSolns},
and making use of well-known results of the theory of ordinary differential
equations ---see e.g. \cite{Har87}--- it follows that, using the same
$\epsilon$ as in the lemma, there exists $\rho_\#>0$ such that for the 
initial data 
\[
\bar{\tau}=0, \quad \bar{\rho}=\rho', \quad  
\psi^A(0)={\psi^A}', \quad \mbox{ with } |\rho'|<\rho_\#,
\] 
and what is implied at these points by  equations \eqref{inx},
\eqref{inX} and \eqref{inf}, the solution of the extended conformal geodesic equations
\begin{equation}
 \bar{\tau}=\bar{\tau}(\tau,\rho',{\psi^A}'), \quad
 \bar{\rho}=\bar{\rho}(\tau,\rho',{\psi^A}'), \quad
 \psi^A=\psi^A(\tau,\rho',{\psi^A}'), \quad \bar{f}_{\bm\alpha}=\bar{f}_{\bm\alpha}(\tau,\rho',{\psi^A}'),
\end{equation}
exist for the values $0\leq\tau\leq1+\epsilon$ of their natural
parameter and the function $\Pi$ is positive in the given range of
$\rho'$ and $\tau$. Moreover, the Jacobian of the map
$(\tau,\rho',{\psi^A}')\rightarrow x^\mu(\tau,\rho',{\psi^A}')$ takes
the value $1+\bar{\tau}$ on $\bar{\cal I}'$ and for sufficiently small
$\rho_\#>0$ it does not vanish in the range $0\leq\tau\leq1+\epsilon$,
$|\bar{\rho}|\leq\rho_\#$. Therefore, the functions
$(\tau,\rho',\psi^A)$ define a smooth coordinate system in a
neighbourhood ${\cal O}'$ of $\bar{\cal I}'$ in $\bar{\cal M}'$. In
particular, the relation \eqref{confF} implies that the curves with
$\rho'>0$ cross ${\scri^+}'$ for $\tau=1$. The set ${\cal O}'$
contains the following special regions
\begin{eqnarray*}
&& {\cal O}'\cap{\scri^+}'=\{\tau=1,\bar{\rho}'>0\}, \\
&&{\cal I}'=\{0\leq\tau<1,\bar{\rho}'=0\}, \\
&&{{\cal I}^+}'=\{\tau=1,\rho'=0\},
\end{eqnarray*}
and it is ruled by conformal geodesics. As a consequence of this
discussion, in terms of the frame fields $v_{\bm \alpha}$ and the
coframe fields $\alpha^{\bm \alpha}$, the
metric $g$, the connection coefficients, $\hat{\Gamma}_{\bm
  \alpha}{}^{\bm \beta}{}_{\bm \gamma}$ of the Weyl connection
$\hat{\nabla}$ and
the components of the tensor fields $\hat{L}_{\bm\alpha\bm\beta}$, $f_{\bm\alpha}$,
$W_{\bm\alpha\bm\beta\bm\gamma\bm\delta}$ extend in the new coordinates as
analytic fields to ${\cal O}'$.

\medskip
Finally, the conformal geodesics on ${\cal O}'$ and the
fields discussed in the previous paragraph can be used to implement
the construction of the manifold $\bar{\cal N}$ as described in
Section \ref{regularProblem}. This is done by solving
linear ordinary
differential equations along the conformal geodesics, with the given
analytical initial data on $\bar{\cal S}$. One obtains the following lemma:

\begin{lemma}
 For stationary asymptotically flat initial data (as described in
Section \ref{stationarySection}) the construction of Section
\ref{regularProblem} leads to a conformal representation of the
stationary vacuum spacetime which, in a neighbourhood ${\cal O\subset \bar{\cal N}}$ of the set
$\bar{\cal I}$,  is real analytic in the radial and
angular coordinates.
\end{lemma}

\subsection{The conformal gauge for the initial data}

The conformal representation discussed in the previous lemma has made
use of  the 3-metric $\breve{h}$ and the conformal factor $\breve{\Omega}$
on $\tc{S}$. It remains to be verified that the whole construction is
robust with respect to rescalings on the conformal initial data of the
form 
\begin{equation}
\label{ChangeConformalGauge}
 \breve{h}\rightarrow\check{h}=\vartheta^2\breve{h}, \quad \breve{\Omega}\rightarrow\check{\Omega}=\vartheta\breve{\Omega},
\end{equation}
where $\vartheta$ is an analytic, positive conformal factor. These
rescalings correspond to a change in the conformal gauge, and imply a harmless change of the normal coordinates
 $x^a\rightarrow {x^a}'$ with ${x^a}'(0)=0$ and an associated change
 $e_{\bm a}\rightarrow e_{\bm a}'$ of the frame vector fields tangent to $\tc{S}$,
 which will be propagated along the new conformal geodesics. It is
 necessary to understand how the congruence of conformal geodesics
 corresponding to $\breve{\Omega}$ relates to the congruence corresponding to $\check{\Omega}$. 

\medskip
The conformal rescaling  given by \eqref{ChangeConformalGauge} also implies the transitions
\[
 \kappa\rightarrow\check{\kappa}=\frac{\vartheta\kappa}{\varsigma},
 \quad \Theta_*\rightarrow\check{\Theta}_*=\varsigma\Theta_*,
\]
where
\[
 \varsigma\equiv \left|1-3\vartheta^{-1}\frac{{\breve{D}}_a\breve{\Omega}{\breve{D}}^a\vartheta}{\Delta_{\breve{h}}\breve{\Omega}}-\frac{3}{2}\vartheta^{-2}\breve{\Omega}\frac{\breve{D}_a\vartheta\breve{D}^a\vartheta}{\Delta_{\breve{h}}\breve{\Omega}}\right|^{1/2}.
\]
The function $\varsigma$ extends to $\bar{\cal S}$ as an analytic
function of $(\bar{\rho},\psi^A)$. From the initial conditions for the
$\breve{\Omega}$-congruence of conformal geodesics one gets that
\[
 \dot{\check{x}}=\varsigma^{-1}\dot{x}, \quad \check{f}_{\tc{S}}=f_{\tc{S}}+\vartheta^{-1}\mbox{d}\vartheta-\varsigma^{-1}d\varsigma,
\]
where the subscripts indicate the pull-back to $\tc{S}$. It can be
verified that 
\[
 \dot{\check{x}}\perp\tc{S}, \quad \check{\Theta}^2\t{g}(\dot{x},\dot{x})=1.
\]
In what follows,  we consider the equations for the
$\check{\Omega}$-congruence in terms of $g$ and its Levi-Civita
connection $\nabla$. As a result of the conformal invariance of
conformal geodesics, it follows that the spacetime curves do not
change (as set points)  writing the equations in this
form. Furthermore, their parameter $\check{\tau}$ remains
unchanged. The $1$-form is transformed according to 
\[
\check{f}\rightarrow
f^*=\check{f}-(\Theta\check{\Theta})^{-1}\mbox{d}(\Theta\check{\Theta}),
\]
and therefore
\[
 \langle f^*,\dot{x}\rangle=0, \quad
 f^*_{\tc{S}}=f_{\tc{S}}+\vartheta^{-1}\mbox{d}\vartheta, \quad \mbox{ on } \tc{S}.
\]
If one expresses the 1-form $f^*$ in terms of the $g$-orthonormal
frame $e_{\bm\alpha}$ satisfying  $e_{\bm0}\perp\tc{S}$, one obtains 
\[
 f^*_{\bm 0}=0, \quad f^*_{\bm a}=f_{\bm a}+\vartheta^{-1}\langle
 \mbox{d}\vartheta,e_{\bm a}\rangle,
 \quad {\bm a}=1,2,3.
\]
The fields $\dot{\check{x}}$ and $f^*_{\bm\alpha}$ are the initial data for
the $\check{\Omega}$-congruence written in terms of $g$,
$e_{\bm\alpha}$ and $\nabla$. As $\varsigma\rightarrow1$ and $\langle
\mbox{d}\vartheta,e_{\bm a}\rangle=O(\bar{\rho})$ as $\bar{\rho}\rightarrow0$, then
\begin{equation}
 \check{\Theta}\rightarrow\Theta,\hspace{1cm}\dot{\check{x}}\rightarrow\dot{x},\hspace{1cm}f^*_{\bm\alpha}\rightarrow f_{\bm\alpha}\hspace{1cm} \mbox{ as } \bar{\rho}\rightarrow 0.
\end{equation}
This means that the limits of the initial data for both congruences coincide on ${{\cal I}^0}'$, and therefore the corresponding curves are identical on ${\bar{\cal I}}'$.

\medskip
Now,  one can go back to the arguments used to show the smoothness of
the construction in terms of the $\breve{\Omega}$-congruence and apply
them to the $\check{\Omega}$-congruence. One concludes that in a
certain neighbourhood ${\cal O}'\subset {\bar{\cal
    M}}'$ of ${\bar{\cal I}}'$  the gauge and construction of Section \ref{regularProblem}
for the $\check{\Omega}$-congruence is as smooth an regular the one
based on the $\breve{\Omega}$-congruence. This result is summarised in
the following lemma:

\begin{lemma}
 For stationary asymptotically flat spacetimes the construction of the
set ${\bar{\cal I}}'$ is independent of the choice of conformal factor
$\Omega$. The set ${\cal I}'$ coincides with the projection
$\pi'({\cal I})$ of the cylinder at space-like infinity as defined in
Section \ref{regularProblem}.
\end{lemma}

This concludes the proof of our Main Theorem ---cfr. Subsection \ref{Subsection:MainResult}.

\section{Conclusions}
\label{Section:Conclusions}
The discussion in the previous section has shown that, for initial data
sets which are stationary in the asymptotic region, the construction
of the cylinder at spatial infinity is as regular as one would expect
it to be. As a consequence, the solutions to the associated regular
initial value problem at spatial infinity are regular at the critical
sets $\mathcal{I}^\pm$ notwithstanding the degeneracy of a subset of the
evolution equations at these sets. As the length of our analysis
shows, this is by no means an obvious result, and it makes evident the
delicate interplay between geometry and properties of differential
equations that the conformal framework allows to resolve. Moreover, it
brings to the forefront the special role played by stationary
solutions in the class of solutions to the Einstein field equations
admitting a smooth compactification at null infinity. 

\medskip
It is worth
pointing out that the analysis carried out in this article is
essentially a spacetime one. A proof of our
main theorem that relies only on properties of stationary data and the
conformal evolution would be, by necessity, much more complicated and
would require an understanding of the structure of the conformal field
equations that is not yet available. 

\medskip
It is expected that our analysis will play an essential role in the
construction of suitable non-time symmetric generalisations of the
rigidity results for asymptotically simple spacetimes in
\cite{Val10a,Val10b}. In this respect, it will also be of interest to
obtain a parametrisation of a large class of initial data sets for
which it is easy to recognise when the data is, in fact,
stationary. This type of characterisations may well require the
consideration of other properties of stationary solutions
at spatial infinity which have not been touched upon here ---most
notably, whether stationary data sets satisfy some generalisation of
the regularity conditions of \cite{Fri98a,Val09b}.

\section*{Acknowledgements}

The authors would like to thank Helmut Friedrich for helpful
discussions on the topic of this article and also on related
matters. Most of this research was done while JAVK was an EPSRC
Advanced Research fellow.

\appendix

\section{Various expansions}\label{expansions}

In this appendix we  collect the expansions of some auxiliary quantities used in this article.

\medskip
Recalling that $\bar{\Omega}=\rho^{-1}V^\frac{1}{2}\Omega$ one obtains 
\begin{eqnarray*}
&& \bar{\Omega} =  \rho+M\rho^2+\frac{1}{6M^2}\rho^3\big(5M^4+2MM_{ab}e^ae^b+2(S_ae^a)^2\big)+O(\rho^4),\\
&& \partial_a\bar{\Omega} =  e_a+2Me_a\rho+\frac{1}{6M^2}\rho^2\big(15M^4e_a+2Me_aM_{bc}e^be^c\\
&& \hspace{2cm}  +4MM_{ab}e^b+2e_a(S_be^b)^2+4S_a(S_be^b)\big)+O(\rho^3),\\
&& \partial_a\partial_b\bar{\Omega} =  \frac{1}{\rho}(\delta_{ab}-e^ae^b)+2M\delta_{ab}+\frac{1}{3M^2}\rho\Big(\tfrac{15}{2}M^4(\delta_{ab}+e_ae_b)\\
&&\hspace{2cm} +M\big(2M_{ab}+4e_{(a}M_{b)c}e^c+(\delta_{ab}-e_ae_b)M_{cd}e^ce^d\big)\\
&&\hspace{2.5cm} +2S_aS_b+4e_{(a}S_{b)}S_ce^c+(\delta_{ab}-e_ae_b)(S_ce^c)^2\Big)+O(\rho^2).
\end{eqnarray*}
Also
\[
 \bar{g}^{ab}\,\partial_a\bar{\Omega}\,\partial_b\bar{\Omega}=-\rho^2-4M\rho^3-\frac{1}{M^2}\rho^4\big(9M^4+2MM_{ab}e^ae^b+2(S_ae^a)^2\big)+O(\rho^5).
\]
The Christoffel symbols of $\bar{g}$ are given by
\[
 \bar{\Gamma}_\mu{}^\rho{}_\nu=\tfrac{1}{2}\bar{g}^{\rho\lambda}\big(\partial_\mu\bar{g}_{\lambda\nu}+\partial_\nu\bar{g}_{\mu\lambda}-\partial_\lambda\bar{g}_{\mu\nu}\big).
\]
It follows that 
\begin{eqnarray*}
&& \bar{\Gamma}_t{}^a{}_t  =  -\frac{1}{2}\bar{g}^{ab}\partial_b\bar{g}_{tt}\\
&& \phantom{\bar{\Gamma}_t{}^a{}_t}= \rho^3e^a+6M\rho^4e^a+\frac{2}{3M^2}\rho^5\big(26M^4e^a+e^aMM_{bc}e^be^c\\
&& \hspace{2cm}+MM^a\,_be^b+e^a(S_be^b)^2+S^aS_be^b\big)+O(\rho^6),\\
&& \bar{\Gamma}_t{}^b{}_a =
\tfrac{1}{2}\big(\bar{g}^{bt}\partial_a\bar{g}_{tt}+\bar{g}^{bc}\partial_a\bar{g}_{tc}-\bar{g}^{bc}\partial_c\bar{g}_{ta}\big)\\
&& \phantom{\bar{\Gamma}_t{}^b{}_a}=  -\rho^3(2S^ce_{ca}\,^b+e^cS^de_{cda}e^b+e_cS_de^{cdb}e_a)+O(\rho^4),\\
&&\bar{\Gamma}_a{}^c{}_b  =  \bar{g}^{ct}\partial_{(a}\bar{g}_{b)t}+\tfrac{1}{2}\bar{g}^{cd}(\partial_{a}\bar{g}_{bd}+\partial_{b}\bar{g}_{ad}-\partial_{d}\bar{g}_{ab})\\
&& \phantom{\bar{\Gamma}_a{}^c{}_b}=  \rho^{-1}(\delta_{ab}e^c-2\delta_{(a}^ce_{b)})+\frac{1}{3M^2}\rho\Big(-M^4\big(\delta_{ab}e^c+e_ae_be^c-2\delta_{(a}^ce_{b)})\\
&&
\hspace{2cm}+2M\big(2e_{(a}M_{b)}\,^c-e^cM_{ab}+2\delta_{(a}^cM_{b)d}e^d-2\delta_{ab}M^c\,_de^d\\
&& \hspace{2cm}+\delta_{ab}e^cM_{de}e^de^e-2e_{(a}e^cM_{b)d}e^d\big)+2\big(2e_{(a}S_{b)}S^c-e^cS_aS_b\\
&& \hspace{2cm}+2\delta_{(a}^cS_{b)}S_de^d-2\delta_{ab}S^cS_de^d+\delta_{ab}e^c(S_de^d)^2-2e_{(a}e^cS_{b)}S_de^d\big)\Big)+O(\rho^2).
\end{eqnarray*}

\end{document}